\documentclass[12pt]{article}
\usepackage{amsmath,amssymb,color}

\newdimen\footheight
\makeatletter
\input sprocl.sty
\makeatother

\usepackage[nosort]{overcite}

\newcommand{\timesw}{\mathrel {\times_{\text{warped}}}}
\newcommand{\AdS}{\operatorname{AdS}}
\renewcommand{\div}{\operatorname{div}}
\newcommand{\Spin}{\operatorname{Spin}}
\newcommand{\SO}{\operatorname{SO}}
\newcommand{\OO}{\operatorname{O}}
\newcommand{\U}{\operatorname{U}}
\newcommand{\SU}{\operatorname{SU}}

\newcommand{\SL}{\operatorname{SL}}
\newcommand{\upperhalf}{\mathfrak h}
\newcommand{\Wpee}{\mathfrak p} 

\newcommand{\Z}{\mathbb Z}
\newcommand{\R}{\mathbb R}
\newcommand{\oo}{\mathfrak o}

\newcommand{\C}{\mathbb C}

\newcommand{\so}{\mathfrak{so}}
\renewcommand{\P}{\mathbb P}
\newcommand{\BL}{\operatorname{Bl}}
\renewcommand{\Im}{\operatorname{Im}}
\newcommand{\su}{\mathfrak{su}}
\renewcommand{\u}{\mathfrak{u}}
\newcommand{\e}{\mathfrak{e}}

\def\ppnumber{\vbox{\baselineskip14pt
\hbox{DUKE-CGTP-01-10}
\hbox{IASSNS-HEP-00/52}
\hbox{NSF-ITP-01-68}
\hbox{hep-th/0411120}
}}

\def\pplogo{\vbox{%
\halign{##&##\hfil\cr&{%
\ppnumber}\cr}
}}
\makeatletter
\def\ps@firstpage{\ps@empty \def\@oddhead{\hss\pplogo}%
  \let\@evenhead\@oddhead 
}
\def\maketitle{
\begin{center}
\quad\vskip.5in
   {\bf \@title \par}       

   \vskip 2em                      
   \@aabuffer\relax
\end{center} \par
\gdef\@aabuffer{}
\thispagestyle{firstpage}
}
\makeatother

\begin{document}

\title{TASI LECTURES ON COMPACTIFICATION AND DUALITY}

\author{ DAVID R. MORRISON }

\address{Center for Geometry and Theoretical Physics,
Duke University, Durham, NC 27708-0318, USA}

\maketitle\abstracts{We describe the moduli spaces of theories with
$32$ or $16$ supercharges, from several points of view.  Included is a
review of backgrounds with D-branes (including type I\,$'$ vacua and
F-theory), a
discussion of holonomy of Riemannian metrics, and an introduction to the
relevant portions of algebraic geometry.  The case of K3 surfaces is
treated in some detail.}

\section*{Introduction}

Although superstring theories themselves are quite restricted in number,
and naturally formulated in ten (spacetime) dimensions, there
is a wide range of possible effective theories in lower dimension which are
obtained by compactifying these theories.  One of the remarkable
features of this story is that such effective theories can often be realized
in more than one way as compactified string theories, a phenomenon referred
to as {\it duality.}  Physical parameters such as the string coupling are
different in the dual descriptions.  Thus, in the parameter space, or
``moduli space,'' for 
the set of theories of a given type, there will be regions where one or
another of the dual descriptions can be studied more easily.  For example,
the effective string coupling may become weak, leading to the
possibility of studying the theory perturbatively in an appropriate limit
(or boundary point) of the moduli space.

In these lectures, we focus on compactifications which have $16$ or $32$
supercharges (a property shared by the original ten-di\-men\-sion\-al
theories). Compactifications with less supersymmetry are treated in Paul
Aspinwall's lectures in this volume.  

We begin with a general discussion of
dualities in lecture~I, working with flat spacetimes only, in which some
dimensions have been compactified into a torus.  One of the surprising
non-per\-tur\-ba\-tive features is the emergence of M-theory, but there are
other interesting dualities as well.  

Not all limiting directions in the
moduli spaces can be studied in this way, so in lecture~II we are led to
introduce D-branes into our superstring backgrounds, applying T-duality to
type~I string theory.  A detailed analysis of the corresponding theories
leads to type~I\,$'$ theory, to F-theory, and to M-theory compactified on a
special type of curved manifold, the K3 surfaces.  

In lecture~III we introduce the general problem of using curved manifolds
as superstring backgrounds, discuss the holonomy classification of
Riemannian manifolds, and are quickly led to introduce the tools of algebraic
geometry for the study of these manifolds.  We give a detailed review of
the relevant portion of algebraic geometry.

Finally, in lecture~IV we return to the case of K3 surfaces and complete
the story of compactifications with $16$ supercharges.

The reader should be familiar with perturbative string theory, as presented
for example in the textbooks of Green, Schwarz, and Witten\cite{GSW}
or of Polchinski.\cite{polchinski}  Good general references for
lectures~I and II are Polchinski's text and a review by
Sen.\cite{SenDuality}  (We also refer the reader to a review by
Mukhi,\cite{Mukhi} and a more comprehensive review by Sen\cite{SenNonp} for
additional details.)  
 A good general reference for lecture~III is the book of Griffiths
and Harris;\cite{griffiths-harris}
in addition, much of the algebraic geometry relevant to string
compactification is 
discussed in the book of Cox and Katz.\cite{CoxKatz}
Further details about K3 surfaces, as discussed in lecture~IV, are
available from either a mathematical\cite{Palaiseau} or
physical\cite{AspinwallReview} perspective. 

\vfill

\section*{Lecture I: S, T, U and all that}

\section{Perturbative superstring theories}

There are five superstring theories.  Each is naturally formulated in ten
dimensions, and can be studied perturbatively at weak string coupling by
means of conformal field theory.  

The five cases are:

\smallskip

\noindent
{\bf Type I.}
A theory of open and closed strings, coupled to
gauge fields taking values in $\so(32)$.
(The global gauge group of the perturbative theory
is\cite{witDK} $\OO(32)/\{\pm1\}$.)
This theory has $16$ supercharges.

\smallskip

\noindent
{\bf Types IIA and IIB.}
A theory of closed strings only (in the perturbative description), with
abelian gauge symmetry in type IIA and no gauge symmetry in type IIB.
The bosonic spectrum includes 
``Neveu--Schwarz-Neveu--Schwarz,'' or NS-NS, fields
(a graviton, a scalar field called
the dilaton, and 
a two-form field),
as well as additional ``Ramond-Ramond,'' or 
R-R,
$p$-form fields,
where $p$ is odd for type IIA and even for type IIB.  
These theories have $32$ supercharges.

\smallskip

\noindent
{\bf Types HE and HO.}
A theory of (heterotic) closed strings only, coupled to gauge fields taking
values 
in $\mathfrak{e}_8\oplus\mathfrak{e}_8$ in type HE, and $\so(32)$ in type
HO. The global gauge groups are  $(E_8\times E_8)\ltimes\mathbb{Z}_2$ for
type HE, and\,\footnote{The notation $\Spin(32)/\Z_2$ denotes a quotient of
$\Spin(32)$ by a
nontrivial $\Z_2$ in the center which does {\it not}\/ yield
$\SO(32)$.  Since the center of $\Spin(32)$ is $\Z_2\times\Z_2$,
there are
two such quotients, but they are isomorphic.} $\Spin(32)/\Z_2$ for type HO. 
The type HE and HO theories each have $16$ supercharges. 

\section{S duality and strong coupling limits}

The duality revolution in string theory began with the realization that
strong-coupling 
limits of the five superstring
theories could be analyzed if certain non-perturbative effects were taken 
into account.  These effects are the result of {\it D-branes}, which are
massive BPS 
states in the  type I and II theories that couple to R-R fields. 
A D$p$-brane is an object in spacetime with $p$ spatial and $1$ time
dimension, on which open strings can end. 
Some D-branes become light at strong coupling, 
where they provide the fundamental degrees of freedom for a dual
formulation of the theory.

We consider the strong coupling behavior case by case.

\smallskip

\noindent
{\bf Type IIB.}
The type IIB theory has two scalar fields: the dilaton, and
the R-R zero-form.  The dilaton couples to the fundamental string of the
theory, while the R-R zero-form couples to the D-string (another
name for the D$1$-brane).  
At strong coupling, this D-string becomes light---the lightest
thing in the spectrum---and exhibits the characteristics of a type IIB
string. 
The conclusion is that the type IIB theory has a weak-strong duality, called
{\it S-duality}.

This conclusion
is further bolstered by  consideration of the type IIB supergravity
action, which describes 
this theory at low energies.  
Letting $\Phi$ denote the dilaton, $G_{\mu\nu}$ the 
graviton
measured in ``string
frame,'' 
$B_2$ the NS-NS two-form field, and $C_p$ the R-R $p$-form field,
the action is invariant under the S-duality transformation
\begin{equation}\label{eq:sdual}
\Phi\mapsto -\Phi,\quad
G_{\mu\nu}\mapsto e^{-\Phi}G_{\mu\nu},\quad
B_2\mapsto C_2,\quad
C_2\mapsto -B_2,\quad
C_4\mapsto C_4
\end{equation}
(setting $C_0=0$ for simplicity).

This symmetry looks more natural if written in  ``Einstein frame''
rather than string frome:
the Einstein frame graviton is
$e^{-\Phi/2}G_{\mu\nu}$, which is 
{\it invariant}\/ under the S-duality transformation given by
Eq.~(\ref{eq:sdual}). 

In string theory, the 
Ramond-Ramond 
fields are invariant under periodic
shifts; in particular, the shift
$C_0\to C_0+1$ leaves the theory invariant.  This combines with the
S-duality to give an $\SL(2,\Z)$ symmetry of the type IIB
theory.\cite{HullTownsend} 

Note that the supergravity action is invariant under
an action of $\SL(2,\R)$.  But the
R-R shifts can only be by integral amounts in string theory, so we expect
precisely $\SL(2,\Z)$ as the symmetry of the type IIB string theory.

\smallskip

\noindent
{\bf Type I.}
There is again a D$1$-brane in this theory which becomes light at strong
coupling. 
However, in this case we see the behavior of a heterotic string in the strong 
coupling limit, rather than type I.  So the weak-strong duality relates two
{\it different}\/ theories: type I and type HO.\cite{witdyn,polwit}

It turns out that non-perturbative effects in type I alter the gauge
group\cite{witDK} from $\OO(32)/\{\pm1\}$  to
$\Spin(32)/\Z_2$, which thus agrees with the (perturbative) gauge group of
type HO theory. This is  explained in John Schwarz's lectures 
in this volume.

\smallskip

\noindent
{\bf Type IIA.}
We get a different behavior this time, due to the light objects being
D-particles, 
i.e., D$0$-branes.  It is believed that there exist bound states of $n$
D$0$-branes 
for every $n$.  Such a bound state will have mass $n/g\sqrt{\alpha'}$.  As
$g\to\infty$, this tower of states
approaches a continuous spectrum whose natural explanation
comes from Kaluza--Klein reduction on a {\it extra}\/ circle of radius
$g\sqrt{\alpha'}$. 
Thus we are led to the conclusion that the strong coupling limit of type IIA
string theory is a mysterious eleven-dimensional theory,
known as ``M-theory''.\cite{townsend,witdyn}
It is not a string theory, but it does have a low-energy description in
terms of 
eleven-dimensional supergravity.
The bosonic field content of M-theory is quite simple, consisting of
a graviton 
and a three-form field.

\smallskip

\noindent
{\bf Types HE and HO.}
We cannot directly analyze the strong coupling limits in these cases with
D-brane 
technology.  However, we can infer from the above discussion that the
strong coupling limit of the type HO string is the type I string. 
We will discuss the strong coupling limit of the type HE string theory in
the next lecture. 

\section{T-duality for type II theories}

Another important duality relating string theories is known as {\it
T-duality}. 
T-duality has a non-trivial effect on the perturbative string, and has been
known for much longer 
than the S-dualities described in the previous section.

T-duality appears when the spacetime on which the string theory is being
formulated 
includes a compact circle $S^1$.  A string wrapped on a circle (or more
generally, 
on a torus $T^d=(S^1)^d$) has {\it winding modes}\/ as well as the
conventional momentum modes.  In the perturbative analysis,
by using a Fourier transform, it can be seen that the 
conformal field theory is
invariant under
\begin{equation}
r\to\alpha'/r,\quad
\text{momentum}\to\text{winding},\quad
\text{winding}\to\text{momentum}.
\end{equation}
(Here, $r$ is the radius of the circle and $\alpha'$ is the string tension.)
This remarkable result relating  
large and small distances was regarded as the first concrete evidence that
string theory must modify our traditional understanding of geometry.

In this section, we discuss T-duality for the type II theories; we shall
return to T-duality in the case of heterotic theories in 
section~\ref{sec:heteroticT}, and in the case of type I theory in lecture~II.

The worldsheet action for strings on a torus depends on a choice of
flat metric on the torus, and a choice of NS-NS two-form field
(the ``B-field'').  We can separate
out the volume as a separate parameter, and recall that the space of
volume-one flat metrics on a torus can be described as
$\SL(d)/\SO(d)$.  The entire parameter space is thus
\begin{equation}
\Gamma_0\backslash\Lambda^2\R^d\times\R^+\times\SL(d)/\SO(d)
\end{equation}
with discrete identifications $\Gamma_0$ coming from two sources:
diffeomorphisms of $T^d$ (which contribute $\SL(d,\Z)$) and integral shifts
of the B-field (which contribute $\Lambda^2\Z^d$).  The total discrete group
coming from this geometrical analysis is
$\Gamma_0=\Lambda^2\Z^d\ltimes\SL(d,\Z)$. 

When T-duality is included, this group becomes much larger: in fact, it
enlarges 
to $\OO(\Lambda^{d,d})$, where $\Lambda^{d,d}$ denotes a lattice with
inner product of signature $(d,d)$, which is {\it even}\/ and {\it
unimodular}. (We are employing standard mathematical terminology here: a
 ``lattice''  has  a bilinear pairing
$\langle\ell_1,\ell_2\rangle$ 
taking values in $\Z$,
 ``even'' means that $\langle\ell,\ell\rangle$ is in $2\Z$ for every
$\ell\in\Lambda$, and
 ``unimodular'' means that for every $\ell_1\in\Lambda$, there is some
$\ell_2\in\Lambda$ such that $\langle\ell_1,\ell_2\rangle=1$.)
It is known\cite{serre}
that $\Lambda^{d,d}$ must be isomorphic to the lattice whose bilinear
pairing has matrix
\begin{equation}
 \begin{pmatrix} 0 & I_d \\ I_d & 0
\end{pmatrix} 
\end{equation}
in an appropriate basis, where $I_d$ is the $d\times d$ identity matrix.

The most
elegant formulation of all of this, essentially due to
Narain,\cite{Narain} exploits 
the isomorphism 
\begin{equation}
\Lambda^2\R^d\times\R^+\times\SL(d)/\SO(d)\cong
\OO(d,d)/\left(\OO(d)\times\OO(d)\right)
\end{equation}
to write the moduli space in the form
\begin{equation}\label{eq:odd}
\OO(\Lambda^{d,d})\backslash\OO(d,d)/\OO(d)\times\OO(d).
\end{equation}

Now we wish to extend this analysis to string theory, going beyond
perturbation theory.  The first remark concerns
the Ramond-Ramond fields: the scalars in our effective
theory which come from the R-R sector essentially transform
in one of the spinor representations of $\oo(d,d)$, in type II theories.
(More precisely, the R-R scalars must be modified by the addition
of some NS-NS and mixed states before they transform in this
way.\cite{BMZ}) 
Moreover, the vectors in a type II theory transform in the {\it other}\/
spinor representation.  Thus, we learn that the appropriate 
symmetry group for the moduli space must be $\Spin(d,d)$ (rather than
$\SO(d,d)$ or some intermediate group), since {\it both}\/
spinor representations must be representations of this group.
This makes a small change in the description of the moduli space, which
should be described as
\begin{equation}
\Spin(\Lambda^{d,d})\backslash\Spin(d,d)/\Spin(d)\times\Spin(d),
\end{equation}
but that actually agrees with the previous description in Eq.~(\ref{eq:odd}).

Moreover, when comparing type IIA and IIB theories, we find that the spinor
representations associated to the R-R scalars and to the vectors
are {\it reversed}\/ by T-duality; that is, the T-duality map
interchanges types IIA and IIB.

There is a potential difficulty in the above analysis when the rest of the
spacetime is not flat, as was recently stressed by Aspinwall and
Plesser.\cite{AspPle} 
One way to think about this difficulty is to notice that we have a
relatively small group $\SL(d,\Z)\rtimes\Lambda^2\Z^d$ and a small number
of T-duality transformations which together generate a specific larger group
$\OO(\Lambda^{d,d})$. 
In order for this to work, many group-theoretical identities involving
$\SL(d,\Z)\rtimes\Lambda^2\Z^d$ and T-dualities must hold.  But if the
moduli spaces in question have less supersymmetry and become subject to
instanton corrections, these identities may fail to hold and the generated
group will be much larger.  This is reminiscent of a familiar phenomenon
when studying symmetries of a quantum field theory:  it may be that the
symmetry algebra can be defined by symmetries which extend off-shell, but
the {\it relations}\/ in the symmetry algebra only hold on-shell.  (One
often says in this situation that the algebra ``closes on-shell.'')
The conclusion is that T-duality holds in the expected form when there
is a large amount of supersymmetry, but in vacua where some of the
supersymmetry is broken, T-duality may also break down.

\section{U-duality}

\begin{figure}[h,t]
\begin{center}
\setlength{\unitlength}{1 true in}
\begin{picture}(1.5,2)(1.65,-.5)
\thicklines
\put(1.9,0){\line(1,0){1}}
\put(1.9,1){\line(1,0){1}}
\put(2.9,1){\line(0,-1){1}}
\put(1.9,1){\line(0,-1){1}}
\put(1.825,-.25){\makebox(.25,.25){\footnotesize $0$}}
\put(2.275,-.25){\makebox(.25,.25){\footnotesize $r_9$}}
\put(2.675,-.25){\makebox(.25,.25){\footnotesize $\infty$}}
\put(1.675,-.05){\makebox(.25,.25){\footnotesize $0$}}
\put(1.675,.45){\makebox(.25,.25){\footnotesize $r_{10}$}}
\put(1.675,.9){\makebox(.25,.25){\footnotesize $\infty$}}
\put(1.575,-.3){\makebox(.25,.25){\footnotesize IIB}}
\put(1.575,1.05){\makebox(.25,.25){\footnotesize IIA}}
\put(2.975,1.05){\makebox(.25,.25){\footnotesize M}}
\put(2.975,-.3){\makebox(.25,.25){\footnotesize IIA}}
\put(1.9,0){\line(1,1){.14}}
\put(2.1,.2){\line(1,1){.14}}
\put(2.3,.4){\line(1,1){.14}}
\put(2.5,.6){\line(1,1){.14}}
\put(2.7,.8){\line(1,1){.14}}
\end{picture}
\caption{Compactifications of M-theory}\label{fig:limits}
\end{center}
\end{figure}
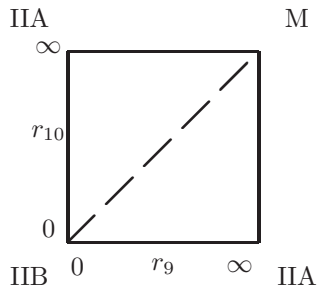

If we put together what we have learned about S-duality and T-duality for
type II theories in nine dimensions, we arrive at the following picture:
starting with M-theory, compactify on $T^2$ with $r_9$, $r_{10}$
being the radii of the circles, and consider limits when $r_j$ gets large
or small (illustrated in Figure~\ref{fig:limits}).

There is a symmetry $r_9\leftrightarrow r_{10}$ which is geometric on
the M-theory side, which generates $\SL(2,\Z)$ on the type IIB side, and
which shows that you get the {\it same}\/ type IIA theory no matter which
M-theory circle you shrink down.\cite{SchwarzFMA,AspinwallFMA}

One feature of this picture which will be important later:
the parameters of the type IIA theory can be written as
\begin{equation}
g_{IIA}=r_{10}^{3/2}, \qquad r_{IIA}=r_{10}^{1/2}r_9
\end{equation}
(measuring the radius in string frame), and the parameters of
the type IIB theory in string frame become:
\begin{equation}
\begin{aligned}
g_{IIB}&=r_{IIA}^{-1}g_{IIA}=(r_{10}/r_9)^{1/2}
\\
r_{IIB}&=1/r_{IIA}=r_{10}^{-1/2}r_9^{-1}
\end{aligned}
\end{equation}
and so in Einstein frame, we get
\begin{equation}
r_{IIB,\text{Einstein}}=g_{IIB}^{-1/2}r_{IIB}=(r_9r_{10})^{-3/4}.
\end{equation}

More generally, we can study the type IIA and IIB theories compactified
on $T^d$ by means of M-theory compactified on $T^{d+1}$.  The
massless bosonic fields in the effective theory are derived
from the graviton and three-form field in eleven dimensions. One thing
which must be specified is a flat metric on $T^{d+1}$ (the expectation
value of the graviton in the compact directions).  These are parameterized
by 
\begin{equation}
\operatorname{Met}(T^{d+1})= \Gamma_0\backslash {\R}^+\times
\SL(d+1,\R)/\SO(d+1),
\end{equation} 
where $\Gamma_0=\SL(d+1,\Z)$ comes from diffeomorphisms of $T^{d+1}$.
The rest of the fields transform in representations of 
$\SL(d+1,\R)$, and we will label them accordingly.

The massless bosonic field content arises from two sources.
As noted above, the M-theory graviton contributes scalars parameterized by
$\operatorname{Met}(T^{d+1})$,
together with $d+1$ vectors and a lower-dimensional
graviton.  On the other hand, 
the M-theory three-form contributes
$\binom{d+1}3$ scalars,
$\binom{d+1}2$ vectors,
$d+1$ two-form fields, and a single three-form field.
The group of discrete identifications is enlarged from 
$\Gamma_0=\SL(d+1,\Z)$ to include a periodicity of $\Lambda^3\Z^{d+1}$ on the
scalars coming from the M-theory three-form.  (As we shall see below, a
further enlargement is in fact expected.)

From the type IIA perspective, the symmetry group $\SL(d,\R)$ is
enhanced to $\OO(d,d)$ through $T$-duality, or to $\SL(d+1,\R)$
through the M-theory interpretation.  Together, these symmetries
generate the larger {\it U-duality group}.  To see what it is,
consider the Dynkin diagram of $\SL(d,\R)$, which is $A_{d-1}$ 
(with $d-1$ nodes):
\begin{center}
\setlength{\unitlength}{1 true in}
\begin{picture}(3.4,.25)(1.4,1)
\thicklines
\put(2.4,1){\circle*{.075}}
\put(2.4,1){\line(1,0){.5}}
\put(2.9,1){\circle*{.075}}
\put(2.9,1){\line(1,0){.5}}
\put(3.4,1){\circle*{.075}}
\put(3.4,1){\line(1,0){.25}}
\put(3.75,1){\circle*{.02}}
\put(3.85,1){\circle*{.02}}
\put(3.95,1){\circle*{.02}}
\put(4.05,1){\line(1,0){.25}}
\put(4.3,1){\circle*{.075}}
\end{picture}
\vskip12pt
\end{center}

\noindent
The enlargement to $\OO(d,d)$ has Dynkin diagram $D_d$:

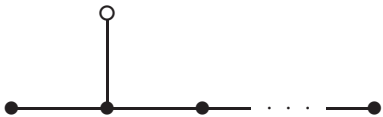
\begin{center}
\vskip12pt
\setlength{\unitlength}{1 true in}
\begin{picture}(3.4,.5)(1.4,1)
\thicklines
\put(2.4,1){\circle*{.075}}
\put(2.4,1){\line(1,0){.5}}
\put(2.9,1){\circle*{.075}}
\put(2.9,1){\line(0,1){.4625}}
\put(2.9,1.5){\circle{.075}}
\put(2.9,1){\line(1,0){.5}}
\put(3.4,1){\circle*{.075}}
\put(3.4,1){\line(1,0){.25}}
\put(3.75,1){\circle*{.02}}
\put(3.85,1){\circle*{.02}}
\put(3.95,1){\circle*{.02}}
\put(4.05,1){\line(1,0){.25}}
\put(4.3,1){\circle*{.075}}
\end{picture}
\vskip12pt
\end{center}

\noindent
and the enlargement to $\SL(d+1,\R)$ is $A_d$ (with $d$ nodes):

\begin{center}
\setlength{\unitlength}{1 true in}
\begin{picture}(3.4,.25)(1.4,1)
\thicklines
\put(1.9,1){\circle{.075}}
\put(1.9375,1){\line(1,0){.4625}}
\put(2.4,1){\circle*{.075}}
\put(2.4,1){\line(1,0){.5}}
\put(2.9,1){\circle*{.075}}
\put(2.9,1){\line(1,0){.5}}
\put(3.4,1){\circle*{.075}}
\put(3.4,1){\line(1,0){.25}}
\put(3.75,1){\circle*{.02}}
\put(3.85,1){\circle*{.02}}
\put(3.95,1){\circle*{.02}}
\put(4.05,1){\line(1,0){.25}}
\put(4.3,1){\circle*{.075}}
\end{picture}
\vskip12pt
\end{center}

\noindent
leading to a combined diagram $E_{d+1}$ (with $d+1$ nodes):

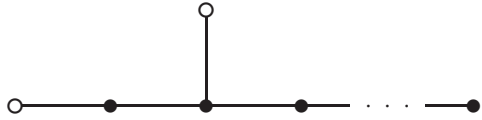
\begin{center}
\vskip12pt
\setlength{\unitlength}{1 true in}
\begin{picture}(3.4,.5)(1.4,1)
\thicklines
\put(1.9,1){\circle{.075}}
\put(1.9375,1){\line(1,0){.4625}}
\put(2.4,1){\circle*{.075}}
\put(2.4,1){\line(1,0){.5}}
\put(2.9,1){\circle*{.075}}
\put(2.9,1){\line(0,1){.4625}}
\put(2.9,1.5){\circle{.075}}
\put(2.9,1){\line(1,0){.5}}
\put(3.4,1){\circle*{.075}}
\put(3.4,1){\line(1,0){.25}}
\put(3.75,1){\circle*{.02}}
\put(3.85,1){\circle*{.02}}
\put(3.95,1){\circle*{.02}}
\put(4.05,1){\line(1,0){.25}}
\put(4.3,1){\circle*{.075}}
\end{picture}
\vskip12pt
\end{center}

\noindent
(Actually, there is another possible combination of $D_d$ and $A_d$,
yielding $D_{d+1}$, if the opposite end of $D_d$ is lengthened, but the
other fields we have which transform under $D_d$ and $A_d$
do so in a way which rules out that combination.)

\begin{table}
\caption{Field content of compactified M-theory}\label{tab:1}
\vspace{0.4cm}
\begin{center}
\begin{tabular}{|c|c|c|c|c|} \hline
&&additional&&two-{}\\
$d{+}1$ & flat metrics on $T^{d+1}$ &{scalars}
&{vectors} & 
forms
\\ \hline
$0$&$\{1\}$&&&\\
$1_A$&$\R^+$&&$\underline{1}$&\underline{1}\\
$1_B$& (Type IIB) & $\underline{2}$ &&$\underline{2}$ \\
$2$ & $\R^+{\times}\SL(2,\R)/\SO(2)$ &  &
 $\underline{2}\oplus\underline{1}$ &$\underline{2}$\\
$3$ & $\R^+{\times}\SL(3,\R)/\SO(3)$ & $\underline{1} 
$ &$\underline{3}\oplus \underline{3}$ &$\underline{3}$\\
$4$ & $\R^+{\times}\SL(4,\R)/\SO(4)$ & $\underline{4} 
$ &$\underline{4}\oplus \underline{6}$ &$\underline{4}\oplus \underline{1}$\\
$5$ & $\R^+{\times}\SL(5,\R)/\SO(5)$ & $\underline{10}
$ &$\underline{5}\oplus \underline{10}\oplus \underline{1}$ &$\underline{5}$\\
$6$ & $\R^+{\times}\SL(6,\R)/\SO(6)$ & $\underline{20} \oplus
\underline{1}$ &$\underline{6}\oplus \underline{15}\oplus \underline{6}$ & \\
$7$ & $\R^+{\times}\SL(7,\R)/\SO(7)$ & $\underline{35} \oplus
\underline{7}$ &$\underline{7}\oplus \underline{21}$ & \\
$8$ & $\R^+{\times}\SL(8,\R)/\SO(8)$ & $\underline{56} \oplus
(\underline{8}\oplus\underline{28})$ & 
&\\ \hline
\end{tabular}
\end{center}
\end{table}

The interpretation of $E_{d+1}$ for small values of $d$ is subtle, and we
have collected all of the necessary data into two Tables.  In 
Table~\ref{tab:1}, we show the various contributions to the scalar and vector
field content of each of the theories,\cite{Julia,DuffLu} and in 
Table~\ref{tab:2} we indicate 
how these are assembled into a symmetric space $G/K$.  Each Table includes
an entry for the type IIB theory in ten dimensions
as well as the various compactifications of M-theory.

The ``additional'' scalar fields come from two sources.  First, as noted
earlier,  the M-theory three-form contributes
$\binom{d+1}3$ scalars.  In addition, 
when the effective dimension $11-(d+1)$ is small, other
 scalars can arise as duals of $p$-form fields.  So when
 $d+1=6$, the three-form dualizes to a scalar, and when $d+1=7$, the
seven two-forms dualize to scalars.  (We shall discuss the case $d+1=8$
momentarily.) 

Similarly, the vectors in the effective theory come from
three sources: Kaluza--Klein vectors, vectors from the three-form,
and ``dual'' vectors (coming from dualizing other fields).  So when
$d+1=5$, the three-form dualizes to a vector, and when $d+1=6$, the six
two-forms dualize to vectors.

In the case $d+1=7$, there are no additional vectors which arise in this
way.  When we go to the case $d+1=8$, {\it all}\/ of the vectors can be
dualized 
to scalars; we do this, and treat both the $8$ Kaluza--Klein vectors and the
$28$ vectors from the three-form as ``additional'' scalars.
All of this is indicated in Table~\ref{tab:1}.

The parameter spaces take the form $\Gamma\backslash G/K$ where
$G{=}E_{d+1(d+1)}$ 
is a noncompact group, $K$ is a maximal compact subgroup,
and $\Gamma$ is a discrete group.  (We have listed the simply-connected
spaces $G/K$ in  Table~\ref{tab:2}.)  Remarkably, the scalars and vectors
assemble themselves into representations of $G=E_{d+1(d+1)}$
in every case.

\begin{table}
\caption{The moduli space of compactified M-theory}\label{tab:2}
\vspace{0.4cm}
\begin{center}
\begin{tabular}{|c|c|} \hline
$d+1$ &  $G/K$\\ \hline
$0$&$\{1\}$\\
$1_A$&$\R^+$\\
$1_B$&  
$\SL(2,\R)/\SO(2)$ 
\\
$2$ & $\R^+{\times}\SL(2,\R)/\SO(2)$ \\
$3$ & $\SL(2,\R){\times}\SL(3,\R)/\SO(2){\times}\SO(3)$ \\
$4$ & $\SL(5,\R)/\SO(5)$ \\
$5$ & $\SO(5,5)/\SO(5){\times}\SO(5)$ \\
$6$ & $E_{6(6)}/\operatorname{Sp}(4)$ \\
$7$ & $E_{7(7)}/\SU(8)$ \\
$8$ & $E_{8(8)}/\SO(16)$ \\
\hline
\end{tabular}
\end{center}
\end{table}

The non-compact groups which are appearing here are so-called ``split
forms.'' 
In general, complex semisimple Lie algebras have a classification by Dynkin
diagrams, and there is a unique connected, simply connected complex group 
for each algebra.  (These are groups like $\SL(n,\C)$ or the universal
cover of $\SO(n,\C)$.)  There are a number of different real groups
whose complexification is a given complex group.  The most familiar ones to 
physicists are the compact groups (such as $\SU(n)$ and $\SO(n)$ in
the examples above.)  But there are also a number of non-compact
groups with the same complexification, such as $\SL(n,\R)$ and
$\SO(p,q)$.  The ``split forms'' are the real groups which are as far
from compact as possible.

The discrete groups $\Gamma$ are believed to be integer versions of
these split groups.\cite{HullTownsend}  The full group $G$ is a symmetry of
the 
compactified supergravity,\cite{CremmerJulia} but string or M-theory should
break this to $\Gamma$.
It is believed that in each dimension other than ten, 
the parameter space for theories with
$32$ supercharges is connected, and is precisely the space $\Gamma\backslash
G/K$ described above. (In dimension ten, the type IIA and type IIB theories
provide different connected components, which we have labeled as $1_A$ and
$1_B$ in Table~\ref{tab:2}.)

\section{Heterotic T-duality} \label{sec:heteroticT}

Returning again to T-duality, we wish to discuss T-duality for heterotic
strings.

The heterotic string theories include gauge fields in the NS-NS sector,
and the Narain analysis is modified by their
presence.  When compactifying on $T^d$, Wilson lines for these gauge fields
are among the parameters.  
We will only consider vacua for which the gauge bundle has trivial topology,
and with the property that
the Wilson lines can be simultaneously conjugated into a 
Cartan torus.  
(This latter property always holds when $d\le2$.)
Imposing these properties leads us to an irreducible component of the
moduli space, called the {\it standard component}, in each dimension less
than ten.  There are many ways to construct other components; we will
discuss these briefly at the end of this section.
  
In ten dimensions, there are two kinds of heterotic theory, and we can
represent the 
Cartan torus of the gauge group from the ten-dimensional theory
in the form $(L_G\otimes\R)/L_G$, where $L_G$ is the root
lattice of the gauge group.\,\footnote{Note that the root lattice
is insensitive to the fact that the gauge group $(E_8\times
E_8)\ltimes\Z_2$ of HE theory is disconnected.}
The two possible root lattices will be denoted $L_{E_8}\oplus L_{E_8}$ and
$L_{(\Spin 32)/\Z_2}$; 
each is an even, unimodular lattice of rank $16$.

Our previous parameter space $O(d,d)/O(d)\times O(d)$ must be supplemented by
$L_G\otimes\R^d$ to include the Wilson lines, with an initial duality group of
\begin{equation}
(\SL(d,\Z)\rtimes\Lambda^2\Z^d)\rtimes L_G\otimes\Z^d
\end{equation}
(suppressing the string coupling).

Again, there is an elegant version of the parameter space essentially due
to Narain:\cite{Narain,NSW}
\begin{equation}
\OO(d,d)/\OO(d)\times\OO(d)\times L_G\otimes \R^d
\cong
\OO(d,d+16)/\OO(d)\times\OO(d+16)
\end{equation}
and when the string coupling is included, there is an additional factor of
$\mathbb{R}^+$. 

As indicated below, T-duality provides identifications between the two
Narain moduli spaces for types HE and HO, and (when applied twice in
succession) 
generates a larger discrete duality group $O(\Lambda^{d,d+16})$.
The gauge group in the effective theory includes both momentum and winding
modes around $T^d$, and its Cartan torus takes the form
$(\Lambda^{d,d+16}\otimes\R)/\Lambda^{d,d+16}$.

It is a useful mathematical fact\cite{serre} that for indefinite
lattices, there is 
only one even unimodular indefinite lattice for each rank and signature
(up to isomorphism).  So our notation $\Lambda^{d,d+16}$ is unambiguous.
Moreover, among {\it definite}\/ lattices, the low rank ones can be
classified: there is only one of rank $8$, namely $L_{E_8}$, and there are
exactly 
two of rank $16$, namely $L_{E_8}\oplus L_{E_8}$ and $L_{\Spin(32)/\Z_2}$.
The theorem guarantees that
\begin{equation}
L_{E_8}\oplus L_{E_8}\oplus\Lambda^{d,d}\cong\Lambda^{d,d+16}
\cong L_{\Spin(32)/\Z_2}\oplus\Lambda^{d,d}
\end{equation}
whenever $d>0$.

Let us consider T-duality in the case $d=1$.  The space
\begin{equation}
\OO(\Lambda^{1,17})\backslash\OO(1,17)/\OO(1)\times\OO(17)
\end{equation}
has exactly two asymptotic boundary points, one associated to the
decomposition $\Lambda^{1,17}\cong L_{E_8}\oplus L_{E_8}\oplus\Lambda^{1,1}$,
and the other to the decomposition $\Lambda^{1,17}
\cong L_{\Spin(32)/\Z_2}\oplus\Lambda^{1,1}$.  
(The string coupling is small in both cases, and we are suppressing it.)
We assign the boundary points the 
interpretations of types HE and HO strings, or large radius and small radius.
T-duality will relate these interpretations.  (See
Polchinski,\cite{polchinski} vol.~2, p.~78 for details.)

In fact, starting from either heterotic theory, there is a simple choice of
Wilson line 
(a group element of order two, in fact) which breaks the gauge algebra 
to\cite{Ginsparg} 
$\so(16)^{\oplus2}\oplus\u(1)^{\oplus2}$.  Globally, the gauge group
becomes\cite{AspMor:new}
$(\Spin(16)^2\times U(1)^2)/\Z_2$ for either theory.  If we leave that
group unbroken, then the only 
remaining parameter is the radius.  An analysis of the massive states shows
that if we map $r\to 1/r$ while exchanging momentum
and winding modes, then the two heterotic theories are exchanged.

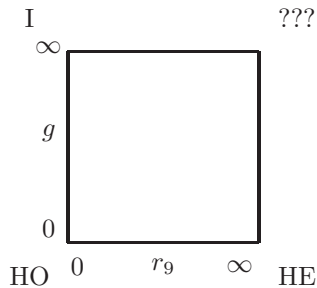
\begin{figure}[h,t]
\begin{center}
\setlength{\unitlength}{1 true in}
\begin{picture}(1.5,2)(1.65,-.5)
\thicklines
\put(1.9,0){\line(1,0){1}}
\put(1.9,1){\line(1,0){1}}
\put(2.9,1){\line(0,-1){1}}
\put(1.9,1){\line(0,-1){1}}
\put(1.825,-.25){\makebox(.25,.25){\footnotesize $0$}}
\put(2.275,-.25){\makebox(.25,.25){\footnotesize $r_9$}}
\put(2.675,-.25){\makebox(.25,.25){\footnotesize $\infty$}}
\put(1.675,-.05){\makebox(.25,.25){\footnotesize $0$}}
\put(1.675,.45){\makebox(.25,.25){\footnotesize $g$}}
\put(1.675,.9){\makebox(.25,.25){\footnotesize $\infty$}}
\put(1.575,-.3){\makebox(.25,.25){\footnotesize HO}}
\put(1.575,1.05){\makebox(.25,.25){\footnotesize I}}
\put(2.975,1.05){\makebox(.25,.25){\footnotesize ???}}
\put(2.975,-.3){\makebox(.25,.25){\footnotesize HE}}
\end{picture}
\caption{Compactifications of Type I and Heterotic  
                                     Strings}\label{fig:hetlimits}
\end{center}
\end{figure}

This leads to a picture in nine dimensions similar to the
one found for the case of $32$ supercharges, and illustrated in 
Figure~\ref{fig:hetlimits}. 

We will investigate the missing corner in the next lecture.  For later
use, we record the relationships among couplings:
\begin{equation}
\begin{aligned}
g_I&=g_{HO}^{-1}, & r_{9,I}&=g_{HO}^{-1/2}r_{9,HO};\\
g_{HE}&=g_{HO}r_{9,HO}^{-1}, &
r_{9,HE}&=r_{9,HO}^{-1}. 
\end{aligned}
\end{equation}
The first line comes from S-duality, and the second line from T-duality.

The string coupling is in fact the only additional parameter in the type
HE and 
HO theories not present in the perturbative analysis, when $d\le4$.  
In that case, the full moduli space is
\begin{equation}
\OO(\Lambda^{d,d+16})\backslash\R^+\times \OO(d,d+16)/\OO(d)\times\OO(d+16)
\end{equation}
(including the string coupling), and
the vectors in the theory transform in the vector representation of
$\OO(d,d+16)$. 
In lower effective, we get further fields in the non-perturbative analysis,
as in the M-theory case: when $d+1=6$ the two-form dualizes to give an
extra vector, when $d+1=7$ the two-form dualizes to give an extra scalar,
and when $d+1=8$ all of the vectors can be dualized to scalars.  As in the
M-theory case, the fields assemble themselves into highly symmetric spaces,
as indicated in Table~\ref{tab:het}.

\begin{table}
\caption{The standard component of the moduli space of compactified
heterotic string theory}\label{tab:het} 
\vspace{0.4cm}
\begin{center}
\begin{tabular}{|c|c|} \hline
$d+1$ &  standard component\\ \hline
$1_E$&$\R^+$\\
$1_O$&   $\R^+$\\
$2$ & $\OO(\Lambda^{1,17})\backslash\R^+\times \OO(1,17)/\OO(1)\times\OO(17)$ \\
$3$ & $\OO(\Lambda^{2,18})\backslash\R^+\times \OO(2,18)/\OO(2)\times\OO(18)$ \\
$4$ & $\OO(\Lambda^{3,19})\backslash\R^+\times \OO(3,19)/\OO(3)\times\OO(19)$ \\
$5$ & $\OO(\Lambda^{4,20})\backslash\R^+\times \OO(4,20)/\OO(4)\times\OO(20)$ \\
$6$ & $\OO(\Lambda^{5,21})\backslash\R^+\times 
   \OO(5,21)/\OO(5)\times\OO(21)$ \\
$7$ & $(\SL(2,\Z)\times\OO(\Lambda^{6,22}))\backslash \upperhalf\times\OO(6,22)/\OO(6)\times\OO(22)$ \\
$8$ & $\OO(\Lambda^{8,24})\backslash \OO(8,24)/\OO(8)\times\OO(24)$ \\
\hline
\end{tabular}
\end{center}
\end{table}

In addition to the standard component, there are many other components of
the moduli space of theories with $16$ supercharges.  For example, there is
a construction known as the CHL string\cite{CHL} which exists in
dimension less 
than ten.  In dimension nine, the CHL string can be described as the HE string
compactified on a circle
with a Wilson line implementing  the $\Z_2$ gauge symmetry 
which exchanges the $E_8$
factors.\cite{CP,LMST}  (This gives a new component in nine dimensions,
since that gauge 
transformation cannot be conjugated into a Cartan torus.)  In eight
dimensions and below, the CHL component can alternatively be described as
the HO string compactified on a circle with a 
non-trivial gauge bundle, the 
bundle ``without vector structure.''\cite{WittenVec}  (These two
descriptions are related by T-duality, as in the case of the standard
component.) The CHL component has
been studied from many points of
view;\cite{SchwarzSen,WittenVec,BerPanSad,BKMT,Mikhailov}
in dimension nine, its moduli space takes the form
\begin{equation}
\OO(\Lambda^{1,9})\backslash\R^+\times \OO(1,9)/\OO(1)\times\OO(9) .
\end{equation}
There are numerous other components in lower dimension, which can be
constructed with a variety of 
different techniques.\cite{CHL,AspinwallFMA,CL,WittenVec,seven}

\vfill

\section*{Lecture II: Backgrounds with Branes}

\section{Type I theory as an orientifold}

The type I string theory can be described as an ``orientifold''
of the type IIB theory.  This means that one introduces the 
orientation-reversal operator $\Omega$ which reverses the
orientation of the worldsheet, and projects to the set of
invariant states, similar to an orbifold projection.
The analogue of the twisted sector in orbifold theory is
provided by new degrees of freedom which can be described as an
{\it orientifold O$9$-plane}\/ together with {\it $16$ D$9$-branes}\/
(projected from $32$ D$9$-branes in type IIB theory).  A collection of $32$
space-filling D$9$-branes in type IIB theory would have $\SU(32)$ gauge
symmetry via the Chan--Paton mechanism, but the Chan--Paton factors
are restricted by the orientifold projection to take values in $\SO(32)$.

In this lecture, we will study models obtained by compactifying type I on a
torus $T^k$ and performing T-duality on $T^k$ (dualizing all compact
directions simultaneously).  As one application of this study,
we will find a weakly coupled dual description of
the strong coupling limit of the 
type HE
string; another application will be to so-called
F-theory models.  For these applications,
we begin with the type HO string compactified on $T^k$, apply S-duality
to get to the type I string,  then apply T-duality and
determine a weakly coupled description of the corresponding theory.

Since the type I theory contains open strings, the T-dual theory will
acquire branes at which the open strings may end---these are just
the standard T-duals of the original D$9$-branes, and give D$p$-branes in the
dual theory (where $p=9-k$).  In addition, the T-dual of the orientifold
operator $\Omega$ is $\Omega$ times a reflection which reverses the T-dualized 
coordinates, i.e., $\iota:(x_1,\dots,x_k)\to(-x_1,\dots,-x_k)$.
There are ``orientifold O$p$-planes'' located at the fixed points of
$T^k/\iota$. 
(We label an orientifold plane according to the number of spatial dimensions
it occupies, just like with a D-brane.  Thus, a D$p$-brane and an O$p$-plane
each occupy $p$ spatial dimensions, i.e., $p+1$ spacetime dimensions.)

The locations of the D$p$-branes in the dual theory are encoded by the
Wilson lines around $T^k$ in the original theory.  So for general Wilson line
values, the dual theory has $2^k$ O$p$-planes and $16$ D$p$-branes deployed
at various locations in $T^k/\iota$.
The O$p$-planes have D$p$-brane charge $-2^{4-k}$ each, so the total D$p$-brane
charge vanishes globally, but the existence of local D$p$-brane charges
means that the Ramond-Ramond fields in the background will not vanish,
but will vary over spacetime.  Such backgrounds are difficult to describe
directly. 

When $k\le4$, we can choose special positions for the D$p$-branes so that
$2^{4-k}$ D$p$-branes are located on top of each O$p$-plane.
In this case, the D$p$-brane charges cancel locally, and no Ramond-Ramond
background is needed; moreover, the model can be studied at weak string
coupling.  The gauge algebra contains a copy of $\so(2^{5-k})$ for each
orientifold plane, so such brane positions must correspond to Wilson lines
in the type I theory which break the nonabelian part of the
perturbative gauge algebra of type I
from $\so(32)$ to $\so(2^{5-k})^{\oplus2^k}$.

It is possible\cite{AspMor:new} to analyze the non-perturbative gauge
groups, for example from the 
heterotic perspective, to obtain the gauge groups given in
Table~\ref{tab:gauge} 
(exploiting the fact that $\Spin(4)\cong\SU(2)^2$).  
For ease of discussion, though, we shall henceforth focus on the gauge {\it
algebras}\/ rather than the global form of the gauge groups.

\begin{table}
\caption{Non-perturbative gauge groups}\label{tab:gauge}
\vspace{0.4cm}
{\renewcommand{\arraystretch}{1.2}
\begin{center}
\begin{tabular}{|c|c|} \hline
$k$ & gauge group \\ \hline
$0$ & $\Spin(32)/\Z_2$ \\
$1$ & $(\Spin(16)^2\times U(1)^2)/\Z_2$ \\
$2$ & $(\Spin(8)^4\times U(1)^4)/\Z_2^3$ \\
$3$ & $(\SU(2)^{16}\times U(1)^6)/\Z_2^5$ \\
\hline
\end{tabular}
\end{center}
}
\end{table}

The descriptions above derive from an analysis of the conformal field theory
(as discussed in Polchinski,\cite{polchinski} for example).  We will, in the
rest of 
this lecture, explore how these models are described in a supergravity
approximation.  

We therefore wish to study {\it backgrounds with branes.}
We model these by using an ansatz for branes which is similar to that
used in studying near-horizon limits (see e.g.\ the lectures by Igor
Klebanov or by Amanda Peet in this volume): for a collection of D$p$-branes, 
we consider a spacetime of the form $\R^{p,1}\times Y^{9-p}$ with
a metric of the form
\begin{equation}
ds^2=H(y)^{-1/2}(-dt^2+dx_1^2+\cdots dx_p^2)+H(y)^{1/2}(g_{ij}dy_idy_j)
\end{equation}
accompanied by a dilaton
\begin{equation}
e^\Phi=H(y)^{(3-p)/4}
\end{equation}
and a Ramond-Ramond field strength
\begin{equation}
F=dt\wedge dx_1\wedge\cdots\wedge dx_p\wedge d(H(y)^{-1}).
\end{equation}
Here, $Y\subseteq \overline{Y}$ is the complement of a finite set of points
$\{P_\alpha\}$ in $\overline{Y}$, $g_{ij}dy_idy_j$ is an appropriate metric on
$Y$ (usually flat or Ricci-flat), and $H(y)$ satisfies
\begin{equation}
\Delta H(y)=\sum_\alpha N_\alpha \delta_{P_\alpha},
\end{equation}
i.e., its Laplacian is a sum of delta functions at the points $P_\alpha$,
weighted by 
integers $N_\alpha$.
Such a metric represents a background with $N_\alpha$ D$p$-branes
located at $P_\alpha$, for each $\alpha$.  (This ansatz must be slightly
modified 
for D$3$-branes, but they will not concern us here.)

\section{Orientifolds in dimension nine}

We begin with the case $k=1$, that is, we analyze the T-dual of the type I
theory 
compactified on $S^1$, choosing the Wilson line to break the gauge algebra to
$\so(16)\oplus\so(16)\oplus\u(1)\oplus\u(1)$.  The dual theory is described
as type IIA on 
$S^1/\Z_2$  
with each endpoint of $S^1/\Z_2$ having an orientifold O$8$-plane and eight
D$8$-branes. (We call this an {\it O$8$ + $8$ D$8$ brane configuration}.)
Since the local D$8$-brane charge is zero, we may use a harmonic function
$H(y)$ on $S^1$ which is $\Z_2$-invariant; since $H(y)$ is harmonic, it must be
{\it linear}.  That is, $H(y)=ay+b$, but then $H(y)=H(-y)$ implies $a=0$ so
$H(y)$ is 
constant.  This leads to a conventional model with constant dilaton and
no Ramond-Ramond flux; however, the space $Y$ is a manifold with boundary, so
the O$8$ + $8$ D$8$ brane configuration
still leaves its mark.

Now we allow the D$8$-branes to move away from the O$8$-planes (which is
accomplished 
on the type I side by allowing the Wilson line to vary).  The function $H$
is now 
only {\it piecewise linear}, and the jumps in its slope measure the jumps in
D$8$-brane charge from region to region in spacetime.\,\footnote{The
piecewise linear nature of the function can be seen directly from a
spacetime analysis,\cite{polwit} or by considering either
D$4$-brane probes\cite{seifive} or D$0$-brane probes.\cite{BSS}}

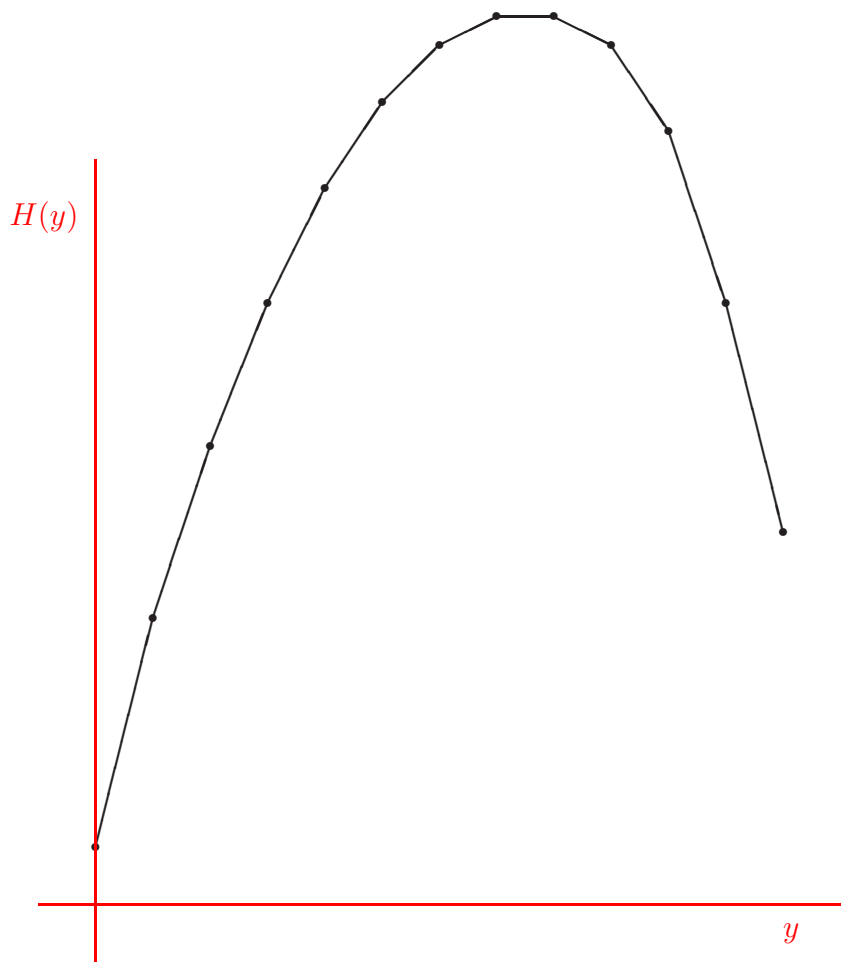
\begin{figure}
\begin{center}
\setlength{\unitlength}{.3 true in}
\begin{picture}(12,16)(0,-1)
\thicklines
\put(0,0){\line(1,4){1}}
\put(1,4){\line(1,3){1}}
\put(2,7){\line(2,5){1}}
\put(3,9.5){\line(1,2){1}}
\put(4,11.5){\line(2,3){1}}
\put(5,13){\line(1,1){1}}
\put(6,14){\line(2,1){1}}
\put(7,14.5){\line(1,0){1}}
\put(8,14.5){\line(2,-1){1}}
\put(9,14){\line(2,-3){1}}
\put(10,12.5){\line(1,-3){1}}
\put(11,9.5){\line(1,-4){1}}

\put(0,0){\circle*{.15}}
\put(1,4){\circle*{.15}}
\put(2,7){\circle*{.15}}
\put(3,9.5){\circle*{.15}}
\put(4,11.5){\circle*{.15}}
\put(5,13){\circle*{.15}}
\put(6,14){\circle*{.15}}
\put(7,14.5){\circle*{.15}}
\put(8,14.5){\circle*{.15}}
\put(9,14){\circle*{.15}}
\put(10,12.5){\circle*{.15}}
\put(11,9.5){\circle*{.15}}
\put(12,5.5){\circle*{.15}}

\color{red}
\put(-1,-1){\line(1,0){14}}
\put(0,-2){\line(0,1){14}}
\put(12,-1.5){\makebox(0,0)[l]{$y$}}
\put(-1.5,11){\makebox(0,0)[l]{$H(y)$}}

\end{picture}
\vskip.25in
\caption{A generic function $H(y)$}\label{fig:1}
\end{center}
\end{figure}

Every function of this type takes the form
\begin{equation}
H(y)=C-\frac12\sum_{i=1}^{16}|y-y_i|
\end{equation}
where $y_i\in[0,1]$ are the locations of the branes, and $y\in[0,1]$ is a 
coordinate on $Y$.  A typical graph of such a function is shown in
Figure~\ref{fig:1}.  Note that the slopes at the endpoints are $\pm 8$,
corresponding 
to the D$8$-brane charge of $-8$ carried by the orientifold O$8$-planes.

The structure near the O$8$-planes can be made more transparent by
extending the function $H(y)$ past those planes, i.e., defining it on $S^1$ 
in a way that is symmetric.  Near $y=0$, this can be done by rewriting
\begin{equation}
H(y)=\widetilde{C}+8|y|-\frac12\sum_{i=1}^{16}(|y-y_i|+|y+y_i|).
\end{equation}
On $S^1$, we have $32$ D$8$-branes in $16$ pairs, related by
reflection.

For generic locations of the D$8$-branes, the gauge symmetry is abelian.
This is enhanced to
$\su(N)$ gauge symmetry when $N$ of the D$8$-branes come 
together (i.e., their $y_i$ values coincide), and to
$\so(2N)$ gauge symmetry 
when $N$ of the D$8$-branes coincide with the O$8$-plane
in an O$8$ + $N$ D$8$ configuration
(i.e., 
$y_i{=}0$ or $y_i{=}1$ 
for all of these).
These gauge symmetry enhancements arise from open strings
stretched between branes which become massless in the limit,
providing off-diagonal elements of a matrix-valued gauge field.

If $N\le7$ D$8$-branes are located at $y=0$, then the initial slope
in $H(y)$ is positive and we can vary the constant $C$ so that 
$H(0)=0$,
without violating the essential requirement $H(y)\ge0$.
There is a further gauge symmetry enhancement at such points\cite{seifive}
from 
$\so(2N)$ to an algebra $\e_{N+1}$.  The new light
particles\cite{KSgauge,MNT,BGL} are D$0$-branes bound to $y=0$,
which are light due to the locally strong coupling at that end 
of $Y$.

A more detailed exploration\cite{delPezzo} of the space of possible
functions $H(y)$ 
reveals additional structure.\,\footnote{Note that in spite of certain
reservations which have been expressed about this picture,\cite{BGS}
the structure of the set of functions $H(y)$ has recently been confirmed
from another point of view.\cite{CachazoVafa}}
There is a phase transition when strong coupling is reached
(i.e., $H(0)=0$ or $H(1)=0$) to another set of models
whose piecewise-linear functions $H(y)$ have $17$ or $18$
singularities.  The slopes at the endpoints of
$[0,1]$ can be as high as $9$, and we interpret the local function
with slope $9$ as
a new kind of spacetime defect: an E$8$-plane.  In
the presence of one or two E$8$-planes, the D$8$-branes are not
free to take arbitrary positions, but are
constrained by the requirement that $H(y)=0$ at the
E$8$-plane(s).

\begin{table}
\caption{Strong coupling limits of orientifold planes}\label{AAA}
\vspace{0.4cm}
\begin{center}
\begin{tabular}{|c|c|c|c|} \hline
$N+1$ & weak coupling & weak coupling & strong coupling\\
   & theory & gauge algebra & theory\\
 \hline
$0$ & -- & -- & E$8$\\
$1_A$& O$8_A$ & $\{0\}$  & (E$8$ + D$8$)$_A$ \\
$1_B$& O$8_B$ & $\{0\}$ & 
(E$8$ + D$8$)$_B$ \\
$2$ & O$8$ + D$8$ & $\so(2)$ & E$8$ + 2 D$8$ \\
$3$ & O$8$ + 2 D$8$ & $\so(4)$ & E$8$ + 3 D$8$ \\
$4$ & O$8$ + 3 D$8$ & $\so(6)$ & E$8$ + 4 D$8$ \\
$5$ & O$8$ + 4 D$8$ & $\so(8)$ & E$8$ + 5 D$8$ \\
$6$ & O$8$ + 5 D$8$ & $\so(10)$ & E$8$ + 6 D$8$ \\
$7$ & O$8$ + 6 D$8$ & $\so(12)$ & E$8$ + 7 D$8$ \\
$8$ & O$8$ + 7 D$8$ & $\so(14)$ & E$8$ + 8 D$8$ \\
\hline
\end{tabular}

\end{center}
\end{table}

A catalog of possible behaviors is given in Tables~\ref{AAA} 
and~\ref{BBB}.  There are some irregularities in the behavior when
the number of D$8$-branes is small:  first, there are two possibilities
for the pure SU(2) probe theory, depending on a 
 $\mathbb{Z}_2$-valued $\theta$-angle; we call the two kinds of
 spacetime defects O$8_A$-planes and O$8_B$-planes.\,\footnote{In
a previous paper,\cite{delPezzo} the O$8_B$-plane was associated with the
``$D_0$ theory'' and the O$8_A$-plane with the ``$\widetilde{D_0}$
theory,'' which 
differ by a $\theta$-angle.\cite{DKV}  Similarly, the (E$8$ + D$8$)$_B$
configuration is associated to the ``$E_1$ theory,'' and the
(E$8$ + D$8$)$_A$
configuration is associated to the ``$\widetilde{E_1}$ theory.''}
Since
 O$8_A$ + D$8$ is physically equivalent to O$8_B$ + D$8$,
 this distinction is
not visible for most orientifold combinations.

Second, the strong coupling limits of the O$8_A$-plane and O$8_B$-plane
give configurations we call (E$8$ + D$8$)$_A$ and (E$8$ + D$8$)$_B$.
A D$8$-brane can be ``emitted'' from an (E$8$ + D$8$)$_A$ configuration
to yield  the exotic E$8$-plane itself.
However, it is not possible to emit a D$8$-brane directly from an
(E$8$ + D$8$)$_B$ configuration.  As in the
O$8$-plane case, 
(E$8$ + D$8$)$_A$ + D$8$
is
physically equivalent to (E$8$ + D$8$)$_B$ + D$8$, so the distinction
is again not visible for most E$8$-plane configurations.  We only
use the $A$ or $B$ subscript on an E$8$-plane in the presence of a single
D$8$-brane. 

The behavior of the function $H(y)$ is different for these two types
of planes: the function for
O$8$ + $N$ D$8$ has a slope of $\pm (8-N)$ at the
boundary, whereas the 
function for E$8$ + $M$ D$8$ has a slope of $\pm (9-M)$ .

Remarkably, the list of strong coupling gauge algebras which appear here
(and which are all compact forms of appropriate ``$\e_{N+1}$'' algebras)
corresponds precisely to the list of U-duality groups in 
lecture I, except that here the {\it compact}\/ algebras appear!

\begin{table}
\caption{Gauge algebras of strong coupling limits, and del Pezzo 
             surfaces}\label{BBB}
\vspace{0.4cm}
\begin{center}
\begin{tabular}{|c|c|c|c|} \hline
$N+1$ &  strong coupling & strong coupling &
   del Pezzo \\ 
   &  theory & gauge algebra & surface \\
 \hline
$0$ & E$8$ & $\{0\}$ & $\P^2$ \\
$1_A$&  (E$8$ + D$8$)$_A$ &
   $\u(1)$ & $\BL_1\P^2$ \\ 
$1_B$&  
(E$8$ + D$8$)$_B$
& $\su(2)$ &
   $\P^1\times\P^1$ \\ 
$2$ &  E$8$ + 2 D$8$ & $\su(2)\oplus \u(1)$ &
   $\BL_2\P^2$ \\ 
$3$ &  E$8$ + 3 D$8$ & $\su(3)\oplus \su(2)$ &
   $\BL_3\P^2$ \\ 
$4$ &  E$8$ + 4 D$8$ & $\su(5)$ & $\BL_4\P^2$ \\
$5$ &  E$8$ + 5 D$8$ & $\so(10)$ & $\BL_5\P^2$ \\
$6$ &  E$8$ + 6 D$8$ & $\e_6$ & $\BL_6\P^2$ \\
$7$ &  E$8$ + 7 D$8$ & $\e_7$ & $\BL_7\P^2$ \\
$8$ &  E$8$ + 8 D$8$ & $\e_8$ & $\BL_8\P^2$ \\
\hline
\end{tabular}

\end{center}
\end{table}

This T-dual of type I theory on $S^1$ is often called type I\,$'$ theory
or type IA theory.  Since it is related to type I by
a T-duality, the couplings are related as
\begin{equation}
g_{I'}=g_Ir_{9,I}^{-1}, \quad r_{9,I'} = r_{9,I}^{-1}.
\end{equation}
We can now analyze the strong coupling limit of the 
type HE
string in nine dimensions.  Combining the
T-duality between types HE and HO with the S-duality between type HO and
type I, we get (as at the end of lecture I)
\begin{equation}
g_I=g_{HE}^{-1}r_{9,HE},\quad r_{9,I}=g_{HE}^{-1/2}r_{9,HE}^{-1/2}
\end{equation}
which is {\it not}\/ a weakly coupled description for strong type HE
coupling (and fixed, large $r_{9,HE}$).  Since the characteristic size
of $r_{9,I}$ is small, we can T-dualize to type I\,$'$:
\begin{equation}
g_{I'}=g_{HE}^{-1/2}r_{9,HE}^{3/2},
\quad r_{9,I'} = g_{HE}^{1/2} r_{9,HE}^{1/2}.\
\end{equation}

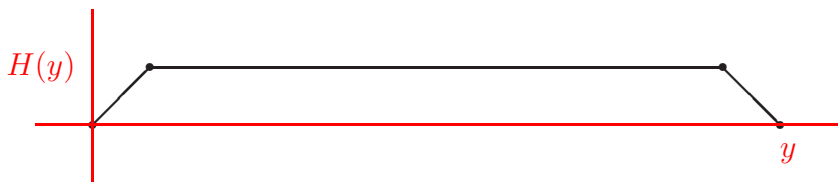
\begin{figure}[h,t]
\begin{center}
\setlength{\unitlength}{.3 true in}
\begin{picture}(12,3)(0,-1)
\thicklines
\put(0,0){\line(1,1){1}}
\put(1,1){\line(1,0){10}}
\put(11,1){\line(1,-1){1}}

\put(0,0){\circle*{.15}}
\put(1,1){\circle*{.15}}
\put(11,1){\circle*{.15}}
\put(12,0){\circle*{.15}}
\color{red}
\put(-1,0){\line(1,0){14}}
\put(0,-1){\line(0,1){3}}
\put(12,-.5){\makebox(0,0)[l]{$y$}}
\put(-1.5,1){\makebox(0,0)[l]{$H(y)$}}

\end{picture}
\caption{The function $H(y)$ for gauge  
                  algebra $\e_8 \oplus \e_8\oplus\u(1)^{\oplus2}$
                  }\label{fig:specialH} 
\end{center}
\end{figure}

These dualities are initially performed with gauge 
algebra
\begin{equation}
\so (16)^{\oplus2}\oplus \u(1)^{\oplus2}
\end{equation} 
which corresponds to $H(y)$ being constant;
we can later tune the Wilson lines to restore
$\e_8\oplus\e_8 \oplus\u(1)^{\oplus2}$
on the heterotic side, and the corresponding tuning on
the type I\,$'$ side yields a function $H(y)$ of the form illustrated in
Figure~\ref{fig:specialH}.

The type I\,$'$ coupling $g_{I'}=H(y)^{-5/4}$ is strong at strong HE
coupling, so we expect a description in terms of M-theory.
We get M-theory scales
\begin{equation}
\begin{aligned}
r_9&=g_{I'}^{-1/3}r_{9,I'} = g_{HE}^{2/3}
\\
r_{10} &= g_{I'}^{-2/3} = g_{HE}^{1/3} r_{9,HE}^{-1}
\end{aligned}
\end{equation}
and so we see that our strong coupling limit decompactifies
the $r_9$ direction.  This leads to the
Ho\v{r}ava--Witten\cite{HoravaWitten} 
picture of the strong coupling limit of the 
type HE
string: it is described by M-theory compactified on $S^1/\Z_2$
with an $E_8$ gauge symmetry group bound to each end of $S^1/\Z_2$.
The exchange of the two ends leads to the additional $\Z_2$ gauge
transformation of this theory.

\section{Orientifolds in dimension eight, and F-theory}

We turn now to the next case:
the T-dual of type I on $T^2$.  As before, we first choose  appropriate
Wilson lines to break the gauge algebra to 
$\so(8)^{\oplus4}\oplus\u(1)^{\oplus4}$, and then
perform a T-duality to obtain a type IIB string on $T^2/\Z_2=S^2$,
with an orientifold O$7$-plane and four D$7$-branes located at each
of the four fixed points of the $\Z_2$ action.  There is no local
D$7$-brane charge near these points, but since the orientifolding 
operator acts as $-1$ on the NS-NS and \hbox{R-R} two-forms, there is
a monodromy 
$\left(\begin{smallmatrix}-1&\hphantom{-}0\\\hphantom{-}0&-1\end{smallmatrix}
\right) \in\SL(2,\Z)$ 
associated with each point.  Moreover, the metric on
$S^2=T^2/\Z_2$ is an orbifold metric, which has  a ``deficit angle''
of $\pi$ at each of the four points.  (In terms of a local coordinate
$z$, the metric takes the form
$|d\sqrt z|^2 = \frac14|z^{-1/2}\,dz|^2$
which gives deficit angle $2\pi\cdot\frac12=\pi$.)

We now wish to vary the Wilson lines on the type I side, and study
the corresponding backgrounds on the type IIB side.  This problem
was analyzed a number of years ago
 (from a slightly different perspective) by Greene, Shapere,
Vafa, and Yau\cite{GSVY}
who showed how to exploit
the $\SL(2,\Z)$ symmetry to produce solutions.  Since we will use
$\SL(2,\Z)$, the resulting backgrounds have no conventional string
theory description: they require that {\it different}\/ $(p,q)$-strings
of the type IIB theory be fundamental at {\it different}\/ points in
spacetime.  Nevertheless, the low energy supergravity
description (with singularities along D$7$-branes) can be analyzed,
and in fact there are other ways to view such theories as limits
of string theories.  Models of this general class are known as
{\it F-theory compactifications.}\cite{VafaF,MVI,MVII,FMW}

Note that in dimension nine we passed from a constant function to a
piecewise-linear function when branes were moved away from the orientifold
planes; here,
we are passing from a constant function to a holomorphic function
that has an $\SL(2,\Z)$-transformation property.

Since we are going to exploit the S-duality of the type IIB string in
constructing these models, we should work in Einstein frame rather
than string frame.  The supergravity description is then in terms
of a metric of the form
\begin{equation}
ds^2=(-dt^2+dx_1^2+\dots+dx_7^2)+H(y_1,y_2)(g_{ij}dy_idy_j)
\end{equation}
with dilaton
\begin{equation}
e^\Phi=H(y_1,y_2)^{-1}
\end{equation}
and R-R field strength
\begin{equation}
F=dt\wedge dx_1\wedge\dots\wedge dx_p \wedge d(H(y_1,y_2)^{-1}).
\end{equation}
The equation of motion for $H(y_1,y_2)$ is
\begin{equation}
\Delta H(y_1,y_2)=\sum_\alpha N_\alpha\delta_{P_\alpha}
\end{equation}
as before.

Introducing a complex coordinate $z=y_1+iy_2$, we treat the harmonic
function 
$H(y_1,y_2)^{-1}$ as the imaginary part of a holomorphic function $\tau(z)$
(away from $P_\alpha$):
\begin{equation}
H(y_1,y_2)^{-1}=\Im \tau(y_1+iy_2)
\end{equation}
where $\tau(z)$ is only well-defined up to $\SL(2,\Z)$
transformations and $\Im \tau(z)>0$ (since the conformal factor $H$ is
always positive).

The Greene--Sha\-pere--Va\-fa--Yau solutions define $H$ in terms
of functions $\tau(z)$
which come from functions on $\upperhalf/\SL(2,\Z)$, where
$\upperhalf=\{\tau:\Im\tau>0\}$ is the upper half plane, 
in order to get finite energy configurations.  In addition to
the complex field $\tau(z)$ with $\SL(2,\Z)$ invariance,
their solutions specify the corresponding Ricci-flat K\"ahler metric as
\begin{equation}
\frac1{2i}\,
\frac{\left|\eta(\tau(z))\right|^4}{\Im\tau(z)}
\, \left|\prod_{\alpha=1}^m(z-z_\alpha)^{-k_\alpha/12}\right|^2
\, dz d\overline{z},
\end{equation}
where
\begin{equation}
\eta(\tau)=e^{2\pi i\tau/24}\prod_n(1-e^{2\pi i\tau n})
\end{equation}
is Dedekind's eta-function.

This metric has a so-called {\it deficit angle}:
in terms of a new variable $\widetilde{z}=z^{1-k_\alpha/12}$,
the metric looks conventional (and flat). But the variable $\widetilde{z}$ does
not traverse a full phase as $z\mapsto e^{2\pi i}z$.
The exponent $k_\alpha/12$, which determines the deficit 
angle%
\ of $2\pi k_\alpha/12$ at
$P_\alpha$, is a function of the type of singularity occurring at
$P_\alpha$.  

In addition to the deficit angle in the metric, 
the function $\tau(z)$ exhibits singular behavior at each singular point:
it is  multi-valued near the singularity, with the multi-valuedness
being given by some fractional linear transformation from $\SL(2,\Z)$,
which describes the change as $z\mapsto e^{2\pi i}z$.

The possible singularities in such functions $\tau(z)$ were 
classified by Kodaira,\cite{Kodaira} and are described in
Table~\ref{tab:Kodaira}. Kodaira's analysis used algebraic geometry, and we
will briefly sketch it in lecture~IV.  The analysis directly produces the
monodromy in each case, as well as providing algebro-geometric data from
which the deficit angle can be determined.  These are both indicated in the
Table. 

\begin{table}
\caption{Kodaira's classification}\label{tab:Kodaira}
\vspace{0.4cm}
{\renewcommand{\arraystretch}{1.1}
\begin{center}
\begin{tabular}{|c|c|c|c|c|} \hline
Kodaira & brane & gauge & 
deficit 
& monodromy \\
notation & configuration & algebra & 
angle 
& \\ \hline
$I_N$, $N>0$ & $N$ D$7$ & $\su(N)$ & ${\displaystyle \frac{N\pi}6}$ & 
$\begin{pmatrix}\hphantom{-}1&\hphantom{-}N\\
  \hphantom{-}0&\hphantom{-}1\end{pmatrix}$\\ 
$I_0^*$ & O$7$ + $4$ D$7$ & $\so(8)$ & ${\displaystyle \pi}$ & 
$\begin{pmatrix}-1&\hphantom{-}0\\\hphantom{-}0&-1\end{pmatrix}$\\
$I_N^*$, $N>0$ & O$7$ + $(N{+}4)$D$7$ & $\so(2N{+}8)$ &  
  ${\displaystyle \pi+\frac{N\pi}6}$ & 
$\begin{pmatrix}-1&\hphantom{-}N\\\hphantom{-}0&-1\end{pmatrix}$\\
$IV^*$ & E$7$ + $6$ D$7$ & $\e_6$ & ${\displaystyle \frac{4\pi}3}$ & 
$\begin{pmatrix}-1&-1\\\hphantom{-}1&\hphantom{-}0\end{pmatrix}$\\
$III^*$ & E$7$ + $7$ D$7$ & $\e_7$ & ${\displaystyle \frac{3\pi}2}$ & 
$\begin{pmatrix}\hphantom{-}0&-1\\\hphantom{-}1&\hphantom{-}0\end{pmatrix}$\\
$II^*$ & E$7$ + $8$ D$7$ & $\e_8$ & ${\displaystyle \frac{5\pi}3}$ & 
$\begin{pmatrix}\hphantom{-}0&-1\\\hphantom{-}1&\hphantom{-}1\end{pmatrix}$\\
$II$ & H$7$  & $\{0\}$ & ${\displaystyle \frac{\pi}3}$ & 
$\begin{pmatrix}\hphantom{-}1&\hphantom{-}1\\-1&\hphantom{-}0\end{pmatrix}$\\
$III$ & H$7$ +  D$7$ & $\su(2)$ & ${\displaystyle \frac{\pi}2}$ & 
$\begin{pmatrix}\hphantom{-}0&\hphantom{-}1\\-1&\hphantom{-}0\end{pmatrix}$\\
$IV$ & H$7$ + $2$ D$7$ & $\su(3)$ & ${\displaystyle \frac{2\pi}3}$ & 
$\begin{pmatrix}\hphantom{-}0&\hphantom{-}1\\-1&-1\end{pmatrix}$\\
\hline
\end{tabular}
\end{center}
}
\end{table}

In addition, we have identified the gauge symmetry for each of Kodaira's
singularity types, and we have attempted to
describe each case as a ``brane configuration'':
the cases of $N$ D$7$ branes and an O$7$ + $(N{+}4)$ D$7$ brane configuration
follow from conventional descriptions of D$7$-branes
and O$7$-planes, with enhanced gauge symmetry determined by
open strings stretching between branes.  There are no solutions
$\tau(z)$ corresponding to an O$7$ + $N$ D$7$ brane configuration with $N<4$.

The E$7$ + $N$ D$7$ brane configurations can be studied as
strong coupling limits of O$7$ + $(N{-}1)$ D$7$ brane configurations,
just as in nine dimensions.  (In this dimension, there are no
$\Z_2$-valued $\theta$-angles to worry about.)
The H$7$ + $N$ D$7$ configurations
are new to eight dimensions, and much less is known about
their explicit description.

In order to get a global solution on $S^2$, we need the
total deficit angle to be $4\pi$.  The generic such solution has
$24$ D$7$-branes, located at distinct points.  In fact,
each of the O$7$ + $4$ D$7$ brane configurations which occurred in our
original
orbifold splits up into six D$7$-branes whose positions
are, however, somewhat constrained.

The function $\tau(z)$ combines the dilaton and the R-R
scalar into a single holomorphic function on $S^2$.
To relate these F-theory compactifications to other
string models, we can further compactify on $S^1$,
remembering that we are working in Einstein frame.
As we saw in lecture I, if $r_{IIB,Einstein}$ denotes
the radius of this compactification, and we compare to M-theory
compactified on $T^2$, we find
\begin{equation}
r_{IIB,Einstein}=(r_9r_{10})^{-3/4}.
\end{equation}
Thus, the small $S^1$-limit will map over to a limit 
in the M-theory model in which
the area of $T^2$ becomes large (and so the supergravity approximation
should be accurate).  Furthermore, we had
\begin{equation}
g_{IIB}=(r_{10}/r_9)^{1/2}
\end{equation}
and more generally, $\tau$ will capture the conformal class of the metric
on $T^2$, which is equivalent to specifying a holomorphic structure.  
There is thus a dual model in 
seven dimensions which takes the form of M-theory compactified
on a four-manifold which is fibered by $T^2$'s with
holomorphic structure dictated by $\tau(z)$.

The four-manifolds of this kind are not unique, but there
{\it is}\/ a unique one for which the holomorphic fibration
has a holomorphic section.  Much evidence has been amassed in favor of a
proposed duality\cite{VafaF,SenF}
\smallskip
\begin{center}
\begin{tabular}{ccccc}
M-theory on & \quad & & \quad & F-theory from \\
elliptically fibered && $\leftrightarrow$ && $\tau(z)$, further\\
manifold with section &&&& compactified on $S^1$\\ 
&&&&(with vanishing Wilson lines)
\end{tabular}
\end{center}
\smallskip
If we use instead one of the four-manifolds for which the holomorphic
fibration does not have a holomorphic section, we find a model in
which a Wilson line along the $S^1$ has been turned
on.\cite{DRMWilson,seven} 

\section{Orientifolds in dimension seven}

Finally, we consider the T-dual of type I compactified
on $T^3$.  We again get a type IIA model on an orbifold $T^3/\Z_2$,
with an O$6$ + $2$~D$6$ brane configuration
located at each fixed point and a gauge algebra
$\so(4)^{\oplus8}\oplus\u(1)^{\oplus6} = 
\su(2)^{\oplus16}\oplus\u(1)^{\oplus6}$.

If we start with the 
type HO
string compactified on $T^3$,
and perform S-duality followed by T-duality, we find the 
following relations among couplings:
\begin{equation}
\begin{aligned}
g_{new}&=\frac{g_{HO}^{1/2}}{r_{7,HO}r_{8,HO}r_{9,HO}}
\\
r_{j,new}&=\frac{g_{HO}^{1/2}}{r_{j,HO}} \quad \text{for } j=7, 8, 9.
\end{aligned}
\end{equation}

Thus, in the strong coupling limit of this type HO compactification,
we are still seeing strong coupling in the type IIA theory, which
suggests an M-theory interpretation.  Lifting the orientifolding
operator to M-theory reverses the sign on the
tenth
spatial dimension, so the dual model is M-theory on $T^4/\Z_2$ with 
$\Z_2$ acting as $-1$ on all four coordinates. There are
$16$ fixed points for this action.

What becomes of the D$6$-branes and O$6$-planes in this
M-theory description?  Our usual ansatz near a D$6$-brane
\begin{equation}
\begin{aligned}
ds^2&=H(y)^{-1/2}(-dt^2+dx_1^2+\cdots+ dx_6^2)
+H(y)^{1/2}(g_{ij}dy_idy_j)
\\
e^\Phi&=H(y)^{-3/4}
\end{aligned}
\end{equation}
should be rewritten in M-theory frame, with metric
$g^{-2/3}ds^2$:
\begin{equation}
d\widetilde{s}^2=(-dt^2+dx_1^2+\cdots+dx_6^2)+\widetilde{g}_{ij}dy_idy_j
+g_{10,10}dx_{10}^2.
\end{equation}
There is no longer a conformal factor in the metric on the worldvolume of
the brane,  
so we simply need a Ricci-flat metric on the remaining four coordinates,
i.e., the D$6$-branes do not appear in this frame.
The general solution with $16$ supercharges gives a K3 surface.

In our original orbifold model, we had $16$ singular points.
Each can be described as a limit of a smooth K3 metric
in which an $S^2$ has shrunk to zero area.  Wrapping
the M-theory membrane around such an $S^2$, we see a new
light particle in the limit, which lies in a vector multiplet
and is responsible for $\su(2)$ enhanced gauge
symmetry.\,\footnote{The story of enhanced gauge symmetry from the point of
view of D$6$-branes is somewhat complicated,\cite{SenGauge}
and we will not discuss it here.}  In this way,
the expected $\su(2)^{\oplus16}\oplus\u(1)^{\oplus6}$ gauge 
algebra is reproduced.

More generally, it is possible to shrink various configurations
of $S^2$'s to get various enhanced gauge
symmetries.  The spacetime singularities always take the
form $\C^2/\Gamma$ for some $\Gamma\subseteq\SU(2)$,
and lead to the possibilities described in Table~\ref{tab:ADE}
(where we list the image of $\Gamma$ in $\SO(3)=\SU(2)/\{\pm1\}$ 
due to the familiar form of the answers).

\begin{table}[h,t]
\caption{Finite subgroups of $\SU(2)/\{\pm1\}$ 
and the associated 
  gauge algebras}\label{tab:ADE}
\vspace{0.4cm}
{\renewcommand{\arraystretch}{1.2}
\begin{center}
\begin{tabular}{|c|c|c|} \hline
image of $\Gamma$ & classical symmetry & gauge \\[-4pt]
in $\SO(3)$ &  & algebra \\
\hline
$\Z_N$ & rotations of $N$-gon & $\su(N)$ \\
$\mathbb D_{2N}$ & rotations and reflections of $N$-gon & $\so(2N)$ \\
$\mathbb T$ & symmetries of tetrahedron & $\e_6$ \\
$\mathbb O$ & symmetries of octahedron or cube & $\e_7$ \\
$\mathbb I$ & symmetries of icosahedron  & $\e_8$ \\[-4pt]
            & or dodecahedron & 
\\ \hline
\end{tabular}
\end{center}
}
\end{table}

\section{Probe-brane theories}

An important verification in each of these cases is 
provided by the study of probe-brane theories.  The branes
we have studied (D$8$, D$7$, D$6$) are large and quite
heavy.  It is possible to introduce a parallel D$(p{-}4)$-brane
into the system as a ``probe'' of the background, treating
the big branes as static and the small branes as
fluctuating.\cite{DouglasGF,BDS,DKPS} 
The worldvolume theories on the probe-branes will have $8$ supercharges
(assuming that the background is otherwise flat).

In the case of D$4$-brane probes of D$8$-branes in type I\,$'$,
this leads to a study of five-dimensional field theories with a
``Coulomb branch'' of dimension one.\cite{seifive,delPezzo,DKV}  These theories
also occur on 
Calabi--Yau threefolds, where (for example) the contraction of
a del Pezzo surface to a point yields the probe theory of an
E$8$ + $n$ D$8$ brane, where $n+1$ is the Picard number of the
del Pezzo surface.

In the case of D$3$-brane probes of D$7$-branes in F-theory,
many four-dimensional field theory phenomena are uncovered,\cite{BDS} such
as the 
behavior of $\SU(2)$ gauge theory with $N_f<4$ matter multiplets (from
analyzing O$7$ + $N_f$ D$7$) 
or the existence of Argyres--Douglas type points (from analyzing H$7$ +
$N_f$ D$7$). 
There are also exotic branes at strong coupling with $\e_n$ gauge
algebras.\cite{DasMuk}

Investigations have also been made of D$2$-brane probes of D$6$-brane
theories \cite{SeibergIR,SeibergWitten3D}
Unfortunately,
time does not permit us to discuss 
any of 
these matters in detail.

\section*{Lecture III: Curved Backgrounds}

\section{Holonomy}

As we saw in the previous lecture, certain natural backgrounds for string
compactification which include D-branes (and yield {\it singular}\/
supergravity 
solutions) break half of the supersymmetry of the original theory.
The other natural way to study models with reduced supersymmetry is to
introduce {\it curved}\/ backgrounds.

The traditional way this has been done in string theory has been to decompose
the ten-dimensional spacetime as a product $X^d\times M^{1,9-d}$ of
a compact manifold $X$ and a flat spacetime $M$.  To understand how much
supersymmetry is preserved in such backgrounds, we must decompose the 
$(9+1)$-dimensional spinor representation according to 
$\Spin(d)\times \Spin(1,9-d)$, and ask how many covariantly constant spinors
will exist on $X^d$ (with respect to the given metric on $X^d$)---these
determine 
the unbroken supersymmetries. (See, for example,
Polchinski,\cite{polchinski} Appendix B.1.)

A variant of this construction is given by the Freund--Rubin
ansatz:\cite{FreundRubin} 
we make a decomposition as a product $Y^{d-1}\times\AdS^{1,10-d}$
together with a nontrivial field strength for one of the supergravity fields.
(Similar constructions can also be made for eleven-dimensional supergravity.)
This time,\cite{DNP} the number of unbroken supersymmetries is determined
by the 
number of {\it Killing spinors}\/ on $Y^{d-1}$.

A more general ansatz which combines both of these ideas is a 
{\it warped product}:  this is a background
of the form $X^d\timesw M^{1,9-d}$ with a metric
\begin{equation}
 ds^2=ds_X^2+\phi(x)ds_M^2
\end{equation}
with an appropriate (Ricci-flat) metric $ds_X^2$, a flat metric $ds_M^2$
on $M$, and a conformal factor $\phi(x)$ depending on $x\in X$,
accompanied by a non-trivial field strength for one of the supergravity
fields.  The
space $X$ should have finite volume, but might not be compact
(due to the presence of branes).

To see why all of these constructions are related, note that 
anti-de Sitter space $\AdS^{1,10-d}$ can be decomposed 
as a warped product of $\mathbb R^+$ and $M^{1,9-d}$.  Then we can rewrite
\begin{equation}
 Y^{d-1}\times\AdS^{1,10-d} = \left(Y^{d-1}\times\mathbb R^+\right)
\timesw M^{1,9-d}
\end{equation}
and the Killing spinors on $Y^{d-1}$ go over to covariantly constant
spinors on $Y^{d-1}\times\mathbb R^+$ (by a theorem of B\"ar\cite{baer}).
When $Y^{d-1}$ is a sphere, we can regard this as a brane solution of
the supergravity theory.
For more general manifolds $Y$, there is an intepretation of this solution
as corresponding to branes at singularities.\cite{MP}

We will mainly focus on the case where $X$ is compact and the spacetime
is an ordinary product (not a warped product).  In this case,
the covariantly constant spinors are determined by the holonomy of the
metric.  (Similarly, the Killing spinors on $Y$ in a Freund--Rubin ansatz
will be determined by the ``Weyl 
holonomy''---but 
the ordinary holonomy is easier to work with.)

\begin{table}
\caption{Irreducible holonomy reps with covariantly constant
spinor}\label{tab:hol1} 
\vspace{0.4cm}
\begin{center}
\begin{tabular}{|c|c|} \hline
Holonomy rep. & Geometry \\ \hline
$\{1\}$ on $\mathbb R$ & flat \\
$\SU(n)$ on $\mathbb R^{2n}$ 
($n\ge3$) 
& Calabi--Yau \\
$\operatorname{Sp}(n)$ on $\mathbb R^{4n}$ & hyper-K\"ahler \\
$G_2$ on $\mathbb R^7$ & $G_2$ \\
$\Spin(7)$ on $\mathbb R^8$ & $\Spin(7)$ \\
\hline
\end{tabular}

\end{center}
\end{table}

The classification of holonomy groups of Riemannian manifolds is given by
the Berger--Simons theorem\cite{Berger,Simons} (see the book of
Besse\cite{Besse} for a complete account of holonomy).  Actually, it is
important to bear in mind that 
there is a holonomy {\it representation}\/ which is being classified,
not just a group.  If we start at a point $x\in X$ and follow a loop which
begins and ends at $x$, parallel transport along that path will transport
tangent vectors at $x$ along to tangent vectors at intermediate points,
finally reaching a tangent vector at $x$ again.  This gives a mapping
from $T_{X,x}$ to itself, and the group generated by all such mappings
is the holonomy group (with $T_{X,x}$ giving the holonomy representation
space).  Note that parallel transport can also be applied to differential
forms and to spinors (in the case of a spin manifold), so once the holonomy
group is known, 
determining the covariantly constant differential forms or spinors is 
a simple exercise in representation theory.  We give the
{\it Berger--Simons classification of irreducible holonomy
representations}\/ in Tables~\ref{tab:hol1} and~\ref{tab:hol2}.
The holonomy representations listed in Table~\ref{tab:hol1} are relevant for
supersymmetric compactification of string theories: each has a covariantly
constant spinor, and each is Ricci flat.  The remaining holonomy
representations (with no covariantly constant spinor) are
listed in Table~\ref{tab:hol2}; these find an application in the study of
moduli spaces of supersymmetric vacua.  (This latter application is
discussed in detail in Paul Aspinwall's lectures in this volume.)

\begin{table}
\caption{Irreducible holonomy reps without covariantly constant
spinor}\label{tab:hol2} 
\vspace{0.4cm}
\begin{center}
\begin{tabular}{|c|c|} \hline
Holonomy rep. & Geometry \\ \hline
$\SO(n)$ on $\mathbb R^n$ & general Riemannian\\ 
  $\U(n)$ on $\mathbb R^{2n}$ & K\"ahler\\
$\operatorname{Sp}(1)\times\operatorname{Sp}(n)/\mathbb Z_2$ 
on
$\mathbb R^{4n}$ & quaternion K\"ahler \\
 $H$ on $\mathfrak{g}/\mathfrak{h}$ & locally symmetric 
space $G/H$
\\ \hline
\end{tabular}

\end{center}
\end{table}

To apply this, we also need to know that every compact Riemannian manifold
with a covariantly constant spinor
admits a finite (unbranched) cover which can be decomposed as a Riemannian
product of a flat torus and a collection of compact Riemannian manifolds
with irreducible holonomy 
representations.
The first step in showing this
is furnished by {\it de~Rham's holonomy
theorem}:\cite{derhamtheorem} If a Riemannian manifold $(M,g)$ is complete,
simply 
connected and if its holonomy representation is reducible, then $(M,g)$ is
a Riemannian product.  (It follows easily that if the original manifold had
a covariantly constant spinor, then so does each factor in the de~Rham
decomposition, and as a consequence the metric on the manifold is Ricci
flat.)   The second step is the {\it Cheeger--Gromoll
theorem}:\cite{CheegerGromoll} If $(M,g)$ is a complete connected Rimannian
manifold with non-negative Ricci curvatuve which admits a line, then
$(M,g)$ is a Riemannian product
$(\overline{M}\times\mathbb{R},\overline{g}\times dt^2)$ where
$(\overline{M},\overline{g})$ is a complete connected Riemannian manifold
with non-negative Ricci curvature, and $dt^2$ is the canonical metric on
$\mathbb{R}$. (Applying this result several times then yields the desired
decomposition.\cite{beauville:varietes}) 

\section{Supersymmetric string compactifications}

If we are interested in compactifications of string theories (or M-theory)
which preserve {\it some}\/ supersymmetry, we should
focus on the flat, Calabi--Yau, hyper-K\"ahler, $G_2$, and $\Spin(7)$ cases.
The last two ``exceptional'' cases are poorly understood, and will not
be discussed further here.  (However, there has been some  
progress in understanding these manifolds---see the recent book of
Joyce\cite{joyce} and references therein.)  The flat case leads to the
study of compact tori, which we have already described in lecture~I.

The Calabi--Yau and hyper-K\"ahler manifolds can be given the following
general characterizations (we assume the manifolds are {\it compact}\/):

{\bf Calabi--Yau manifolds} 
(holonomy $\SU(n)$, $n\ge3$) have
a non-vanishing holomorphic $n$-form $\Omega$ and a K\"ahler metric.
There is a unique complex structure (up to complex conjugation)
compatible with the metric.  The K\"ahler metric can be described in
terms of the {\it K\"ahler form}\/
$\omega=\frac i2\sum g_{z_i\bar{z}_j}dz_i\wedge d\bar{z}_j$.
($\Omega$ has a local description of the form
$f(z)dz_1\wedge\cdots\wedge dz_n$ with $f(z)$ holomorphic.)

{\bf Hyper-K\"ahler manifolds} 
(holonomy $\operatorname{Sp}(n)$) have real
dimension $4n$, with a distinguished three-plane of two-forms,
and an $S^2$ of compatible complex structures.  If we choose one
of the complex structures, there is a holomorphic two-form of
the form $\omega_1+i\omega_2$ and a K\"ahler form $r\omega_3$
for some orthogonal basis $\omega_1, \omega_2, \omega_3$ of
the three-plane, and some positive constant $r$.  The manifold
 also has a holomorphic $4$-form, $6$-form, \dots, $2n$-form
 given by taking powers of $\omega_1+i\omega_2$.  In particular,
 there is a form
 $\Omega=(\omega_1+i\omega_2)^n$ of top degree.  It is non-vanishing.
 
The metrics in all of these cases are Ricci-flat.  Such metrics
were studied by Calabi in the 1950's who showed that for a given
complex structure and de Rham cohomology class of K\"ahler metrics,
there is at most one Ricci-flat metric in the class.  (That is,
if $\omega$ is the K\"ahler form of a Ricci-flat metric, then there
is no one-form $\eta$ on $X$ such that $\omega+d\eta$ is also the 
K\"ahler form of a Ricci-flat metric.)
 
Calabi conjectured\cite{calabi} the existence of such metrics, and this
was proved 
by Yau\cite{yau} in the 1970's in the following form:
given a compact complex manifold $X$ of complex dimension $n$ which admits
a non-vanishing holomorphic $n$-form $\Omega$, and given a K\"ahler
form $\omega$ on $X$, there exists a Ricci-flat metric on $X$ whose
K\"ahler form is in the same de Rham class as $\omega$, and for which
$\Omega$ is covariantly constant.

The proof is a non-constructive existence proof.  In particular, although
we are certain that these metrics exist, it is very difficult to calculate
any of their properties.
 
However, this theorem is very powerful as a tool for studying string
backgrounds, since it reduces the search for solutions to the supergravity
equations of motion to a search for complex K\"ahler manifolds which have
a non-vanishing holomorphic $n$-form $\Omega$.
 
In fact, the search can be restricted even further:\cite{beauville:varietes}
it turns out that
for every compact $\SU(n)$ holonomy manifold ($n\ge3$), the complex
structure is {\it algebraic}\/ (i.e., $X$ comes from algebraic geometry);
for hyper-K\"ahler manifolds, generically if you fix the Ricci-flat
metric there will be choices out of the $S^2$ of complex structures for
which $X$ is algebraic.
 
So we can restrict our search to algebraic geometry, and employ a completely
different set of tools to find and study such objects.

\section{Algebraic geometry: a brief introduction}
 
The ``algebraic varieties'' we now must study are complex submanifolds $X$
of complex projective space $\mathbb P^N$.  We describe $\mathbb P^N$
by means of ``homogeneous coordinates'' $[z_1,z_2,\dots,z_N]\ne[0,0,\dots,0]$
which do not label points uniquely but are subject to identifications
\begin{equation}\label{eq:idents}
 [z_0,z_1\dots,z_N]=[\lambda z_0,\lambda z_1,\dots,\lambda z_N] 
\end{equation}
for non-zero complex numbers $\lambda$.  (We use square brackets to 
emphasize that these are not ordinary coordinates.)
 
Given $X\subseteq\mathbb P^N$, each homogeneous coordinate $z_i$ determines
a codimension one subvariety
\begin{equation}
D_i=X\cap \{z_i=0\} 
\end{equation}
on $X$.  (We are assuming that $X\not\subseteq\{z_i=0\}$; otherwise,
we would have treated $X$ as a submanifold of $\{z_i=0\}\cong\mathbb
P^{N-1}$.) 
Such a codimension one subvariety is called an
{\it effective divisor}\/ on $X$.
More generally, a combination $\sum m_iD_i$ of effective divisors
with integer coefficients is
called a {\it divisor}.
 
If we consider two of these divisors, $D_i$ and $D_j$, the ratio $z_i/z_j$
makes sense as a function on $X-D_i-D_j$.  (The individual homogeneous 
coordinates are not functions on $X$ or on $\mathbb P^N$ due to the
identifications in Eq.~(\eqref{eq:idents}), but the identifications cancel
out in ratios.)  This 
ratio $z_i/z_j$ extends to a {\it meromorphic function}\/ on $X$: its
only singularities are poles.
 
Generally, for a meromorphic function $f$ defined on $X$, we define the
{\it divisor of $f$}\/ to be
\begin{equation} 
\div(f)=\{f=0\}-\{f=\infty\} 
\end{equation}
where $\{f=0\}$ and $\{f=\infty\}$ are codimension one in $X$.
In the example at hand, we have
\begin{equation} 
\div(z_i/z_j)=D_i-D_j .
\end{equation}
This property is characteristic of divisors which occur as intersections
with linear functions in $\mathbb P^N$ for the {\it same}\/ embedding
in projective space.  In general, a given algebraic variety will have
many different embeddings into projective spaces.
 
To determine all of the ways to embed $X$ into projective spaces, we
can study {\it all}\/ of the divisors on $X$.  To determine which divisors belong
to the same embedding, we introduce some definitions. Two divisors $D$ and $D'$
are said to be {\it linearly equivalent}\/ if there is a meromorphic 
function $f$ such that 
\begin{equation} 
\div(f)=D-D' .
\end{equation}
The {\it linear system containing $D$}\/ is the set
\begin{equation}
\begin{aligned}
|D| = \{ D' \ |\  &\text{$D'$ is linearly equivalent to $D$,
and $\textstyle D'=\sum n_iD_i$}\\
&\text{with $n_i\ge0$ and $D_i\subseteq X$ codimension one}\}.
\end{aligned} 
\end{equation}
The last requirement in the definition come from the observation that the
divisors we encountered from $X\subseteq \mathbb P^N$ were effective divisors,
i.e.,
subsets of $X$ counted with multiplicity, but
with no negative coefficients allowed.

Given a linear system $|D|$, we choose a basis $D_0=D$, $D_1$, \dots, $D_n$
of the divisors in $|D|$, and let $f_1$, $f_2$, \dots, $f_n$ be the
meromorphic functions satisfying
\begin{equation} 
\div(f_j)=D_j-D_0 .
\end{equation}
Then we can define a mapping $X\to\mathbb P^n$ by
\begin{equation}\label{eq:map}
x\mapsto[1,f_1(x),f_2(x),\dots,f_n(x)].
\end{equation}
This is ill-defined along $D_0$, but by exploiting the equivalence in
$\mathbb P^n$ 
we can rewrite this as
\begin{equation} 
[1,f_1(x),f_2(x),\dots,f_n(x)] = [{\textstyle\frac1{f_1(x)}}, 1,
{\textstyle\frac{f_2(x)}{f_1(x)}},\dots,{\textstyle\frac{f_n(x)}{f_1(x)}}] 
\end{equation}
which is ill-defined along $D_1$ instead of along $D_0$, and so on,
for other divisors $D_j$.

Thus, if the divisors $D_0$, $D_1$, \dots $D_n$ have no points in common,
our prescription Eq.(~\ref{eq:map}) can be extended to a well-defined mapping
on all of $X$.  In this case, $|D|$ is said to be {\it base point free.}

The linear system $|D|$ is said to be {\it very ample}\/\,\footnote{A linear
system $|D|$ is 
{\it ample}\/ if some positive multiple $|mD|$ is very ample.} if the
associated 
mapping is actually an embedding into $\mathbb P^n$.  Given a very ample linear
system $|D|$, i.e., an embedding $X\subseteq\mathbb P^n$, we get a natural
K\"ahler metric on $X$ by restricting the Fubini--Study metric from
$\mathbb P^n$. 
Explicitly, the K\"ahler form of this metric on $\mathbb P^n$ can be written
$\omega=\partial\bar\partial \log \sum |z_i|^2$.  Restricting to $X$, we
get a form $\omega_{|D|}=\left(\partial\bar\partial \log \sum
|z_i|^2\right)|_X$ 
 on $X$.

A key fact is that for Calabi--Yau manifolds, the K\"ahler classes
$\omega_{|D|}$ 
coming from projective embeddings will generate all K\"ahler classes (using
positive real linear combinations).\,\footnote{This is because
$h^{2,0}=0$, so the positive rational linear combinations of very ample
classes will be dense in the K\"ahler cone.}  So this portion of our
problem---determining 
the set of K\"ahler classes---can be solved using algebraic geometry.
(The hyper-K\"ahler case is different and will be discussed in lecture IV.)

The other portion of our problem---determining the set of complex
structures---is 
also a problem in algebraic geometry.  Once $X$ has been embedded in 
$\mathbb P^n$, it can always be described by means of a finite set of
homogeneous 
equations 
\begin{equation}
f_1(z_0,\dots,z_n), \dots, f_k(z_0,\dots,z_n),
\end{equation} 
with
\begin{equation} 
X=\{[z_0,\dots,z_n]\in\mathbb P^n\ |\ f_j(z_0,\dots,z_n)=0 
\ \text{for all $j$}\}.
\end{equation}
In principle, the other complex structures are found by varying the
coefficients  
in these defining equations.  There are two difficulties with this in
practice:
\begin{enumerate}
\item There may be some complex structures on this manifold with don't
embed into the same projective space as $X$.
\item The number of equations needed to describe $X$ is larger than
$\dim\mathbb P^n-\dim X$, and the equations don't meet transversally;
thus, if we vary the coefficients arbitrarily we will find a common
intersection 
which is smaller than $X$.  (So we must vary the coefficients {\it
judiciously}, 
and it is hard to see explicitly how to do this.)
\end{enumerate}
We will encounter both of these phenomena in our discussion of K3 surfaces 
below.  Over the years, algebraic geometers have developed some rather
sophisticated machinery to address these issues.  
(See, for example, the treatise of Viehweg.\cite{viehweg})
Very little of this
machinery has been applied to cases of interest in physics (to date!).

So we have seen that the complex structures can be studied by varying
coefficients, and the K\"ahler classes can be studied by locating all
(very ample) divisors.  The issue we have not yet addressed is: how can we
recognize whether or not there exists a non-vanishing holomorphic $n$-form?

A very useful tool in studying this issue is the ``adjunction formula.''
Given a complex submanifold $D\subseteq X$ defined by a single equation
$f=0$ (locally), there is a ``Poincar\'e residue formula'' relating
meromorphic $n$-forms on $X$ and meromorphic $(n-1)$-forms on $D$: 
given a meromorphic $n$-form
\begin{equation} 
\frac{g(w_1,\dots,w_n)dw_1\wedge\cdots\wedge dw_n}{f(w_1,\dots,w_n)} 
\end{equation}
with a simple pole on $D$ (using local coordinates $w_1,\dots,w_n$
on $X$), its Poincar\'e residue is
\begin{equation} 
\left.\frac{g(w)dw_1\wedge\cdots\wedge dw_{n-1}}{\partial f/\partial w_n}
\right|_D 
\end{equation}
(which is well-defined if $\partial f/\partial w_n\ne0$) with similar,
equivalent, 
formulas when $\partial f/\partial w_j\ne0$.  (If $D$ is a
sub{\it manifold}, then at every point one of the $\partial f/\partial w_j$'s
must be ${}\ne0$.)

It is common to express the properties of meromorphic $n$-forms in terms
of {\it divisors}; if $\alpha(w_1,\dots,w_n)dw_1\wedge\cdots\wedge dw_n$
is a meromorphic $n$-form, we define the {\it canonical divisor of $X$}\/
to be
\begin{equation} 
K_X=\div(\alpha)=\{\alpha=0\}-\{\alpha=\infty\}.
\end{equation}
Thus, in our Poincar\'e residue formula, we see
\begin{equation}
K_X=\div(g)-\div(f)=\div(g)-D
\end{equation}
while
\begin{equation}
K_D=\div(g)|_D
\end{equation}
(since $\partial f/\partial w_j\ne0$).  Thus,
\begin{equation}
K_D=(K_X+D)|_D.
\end{equation}
This is known as the {\it adjunction formula}.

The interpretation of $D|_D$ is this: find a divisor $D'$ which is
linearly equivalent to $D$, and treat $D'|_D$ as a divisor on $D$.
(All facts about these divisors are being considered up to linear
equivalence only.)  This is the divisorial version of the
``normal bundle'' of $D$.

The requirement in Yau's theorem is that
there exists a meromorphic $n$-form whose divisor is trivial,
i.e., it has neither zeros nor poles; this is the same as saying that
$K_X=0$. 

{\bf Key example.}  
$K_X=0$ and $D\subseteq X$ is a codimension
one submanifold.  The adjunction formula tells us that $K_D=D|_D$.
Note that $D|_D$ becomes quite concrete if we embed $X$ in $\mathbb P^n$
using the linear system $|D|$: then $D'|_D$ represents the intersection
of $D$ with some $z_i=0$.  In other words, $D$ is embedded by the
canonical linear system $|K_D|$.

\section{Algebraic geometry of K3 surfaces}

The case of $\dim_{\mathbb{C}} X=2$ (an algebraic K3 surface) is instructive.  
In this case, $D$ is a
Riemann surface, which must have some genus $g$.  The degree of the
canonical divisor of a Riemann surface is well-known: $\deg(K_D)=2g-2$.
Also, the canonical linear system $|K_D|$ embeds $D$ into
$\mathbb P^{g-1}$.

The interpretation of these facts in terms of $X$ is that $X$ should
embed in $\mathbb P^g$, and its {\it degree}\/ (the number of
points of intersection $X\cap\{z_i=0\}\cap\{z_j=0\}$) should be
$2g-2$.

Remarkably, surfaces $X\subseteq\mathbb P^g$ of this type exist
for every $g\ge2$, and in every case, the number of independent
deformations of complex structure is $19$.  These are the
{\it algebraic K3 surfaces}.

Let us consider these surfaces for low values of $g$.
\begin{itemize}
\item[\underline{$g=2$}] Riemann surfaces of genus two are hyperelliptic,
and map two-to-one onto $\mathbb P^1$.  So $X$ will map two-to-one onto
$\mathbb P^2$.  The map on $D$ must have six branch points (in order to get
genus two), 
so the map $X\to\mathbb P^2$ must be branched over a curve of degree six.
We can describe $X$ by an equation of the form
\begin{equation}
y^2=z_0^6+z_1^6+z_2^6+\cdots
\end{equation}
(the degree six equation on the right hand side can be arbitrary), and
regard this as an 
equation in a {\it weighted projective space}\/ $\mathbb P^{1,1,1,3}$
in which
$[z_0,z_1,z_2,y]=[\lambda z_0,\lambda z_1,\lambda z_2,\lambda^3 y]$.
(The superscripts in the notation denote the powers of $\lambda$, the
so-called {\it weights}\/ of the homogeneous variables.)
\item[\underline{$g=3$}] The general Riemann surface of genus three
embeds as a degree four curve in $\mathbb P^2$; $X$ should be a 
surface of degree four in $\mathbb P^3$, for example,
\begin{equation}
z_0^4+z_1^4+z_2^4+z_3^4=0.
\end{equation}

\item[\underline{$g=4$}] This time, $D\subseteq\mathbb P^3$ is the
intersection of surfaces of degrees two and three, and $X\subset\mathbb P^4$
will be the intersection of hypersurfaces of degrees two and three.
The degree is then $2\cdot 3=6$.
\item[\underline{$g=5$}] $D\subseteq \mathbb P^4$ and $X\subseteq\mathbb P^5$
can be described as the intersection of three hypersurfaces of degree two.
The degree is $2\cdot2\cdot2=8$.
\item[\underline{$g\ge6$}] $D\subseteq \mathbb P^{g-1}$ and $X\subseteq
\mathbb P^g$ 
require more defining equations than their codimension.  This makes the moduli
problem tricky---as indicated above, coefficients must be varied {\it
judiciously}. 

\end{itemize}
One interesting feature to note about this set of examples:
the complex dimension of the space of Riemann surfaces of genus $g$
is $3g-3$, whereas the dimension of those which lie on a K3 surface is
at most $19+g$ ($19$ parameters for $X$ and $g$ parameters for the choice
of $D$ when $X\subseteq \mathbb P^g$).  Thus, when $g>11$, not every
curve lies on a K3 surface.

Another interesting feature, to be discussed further in lecture IV,
is that the set of all complex structures on a K3 surface has complex
dimension $20$, and form a single family containing all of the algebraic
K3 surfaces of every $g$.  This is an example of the phenomenon mentioned
above in which not all deformations of complex structure may happen
in the given projective space.

However, when the holonomy is $\SU(n)$, $n\ge3$, there always exist
embeddings $X\subseteq\mathbb P^N$ for which all nearby complex
structures can be obtained within the same $\mathbb P^N$.  (Warning:
if somebody hands you $X\subseteq\mathbb P^N$, it might not have this
property:   
some projective embeddings are ``deficient'' in this sense.)

\section{Calabi--Yau manifolds in higher dimension}

The theory of K3 surfaces is understood in great detail, and will
be explained further in the next lecture.
We know much less about the algebraic geometry of Calabi--Yau or
hyper-K\"ahler manifolds 
of higher dimension.  There are two strategies which might be
followed:
\begin{enumerate}
\item try to directly generalize constructions like the $g\le5$ cases
of K3 surfaces
\item try to study in general the possible divisors $D$ and whether
they occur on Calabi--Yau or hyper-K\"ahler manifolds.
\end{enumerate}

The first strategy has led to an extensive study of Calabi--Yau
``complete intersections'' in projective spaces, and more generally
in weighted projective spaces or toric varieties (a further 
generalization of weighted projective space).  At least tens of thousands
of examples have been produced in this way.\,\footnote{For example, there are
473,800,776 types of Calabi--Yau hypersurfaces in four-dimensional toric
varieties, which give at least 30,108 distinct examples, based on Hodge
numbers.\cite{KreuzerSkarke}} 

  And yet, as the above
story about K3 surfaces illustrates, such constructions may have
only barely scratched the surface.

It is instructive to see why the set of complete intersection
Calabi--Yau manifolds (of fixed dimension) in projective space
is finite.  Suppose $X^d\subseteq\mathbb P^n$ has been defined
as the intersection
\begin{equation}
X=Y_1\cap Y_2\cap \cdots \cap Y_{n-d}
\end{equation}
of $n-d$ hypersurfaces.  Each $Y_j$ is linearly equivalent to
$m_jH$, where $H=\{z_0=0\}$, and $m_j$ is the degree of the
homogeneous polynomial defining $Y_j$.  We use that fact that
$K_{\mathbb P^n}=-(n+1)H$ (which can be seen from the existence of a
globally well-defined meromorphic $n$-form
\begin{equation}
\frac{z_0dz_1\wedge dz_2\wedge\cdots\wedge dz_n
+ z_1dz_2\wedge \cdots \wedge dz_n\wedge dz_0 + \cdots}
{z_0z_2\cdots z_{n-1}z_n} 
\end{equation}
with poles along all $n+1$ coordinate hyperplanes)
and apply the adjunction formula repeatedly:
\begin{equation}
\begin{aligned}
K_X=&(\cdots((K_{\mathbb P^n}+Y_1)|_{Y_1}+Y_2)|_{Y_2}+
\cdots+Y_{n-d})|_{Y_{n-d}}\\
&=\left.\left(-(n+1)H+m_1H+m_2H+\cdots+m_{n-d}H\right)\right|_X .
\end{aligned} 
\end{equation}
So the condition to get a Calabi--Yau or hyper-K\"ahler manifold
is: $\sum_{j=1}^{n-d}m_j=n+1$.
Moreover we can choose $m_j\ge2$ for each $j$ since otherwise
$X$ would sit in a linear subspace (a smaller $\mathbb P^n$).
So the condition can be rewritten as
\begin{equation}
\sum_{j=1}^{n-d}(m_j-1)=d+1, \quad m_j-1\ge1 
\end{equation}
and with a fixed $d$ there are clearly only a finite number
of solutions.  The K3 examples described above are reproduced by the solutions
$3=3$, $3=2+1$, $3=1+1+1$, giving degrees
$4$, $(3,2)$, and $(2,2,2)$, respectively, with hyperplane sections having
 $g=3$, $g=4$, and $g=5$.

The second strategy would seem to be a more general one:
first, we study all surfaces $D$ for which $|K_D|$ gives
an embedding (or at least a reasonable map, like the two-to-one map
we encountered for Riemann surfaces of genus two), and then we
try to decide which ones can be on Calabi--Yau threefolds.
As the remark about not all curves lying on K3 surfaces indicated, the
second part will be highly non-trivial.  However, even the first part
is quite hard.

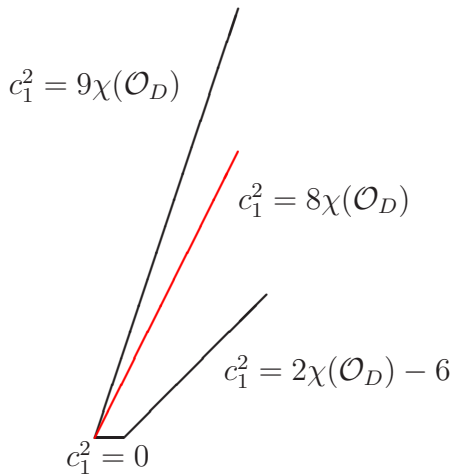
\begin{figure}[h,t]
\vspace{0.6cm}
\begin{center}
\setlength{\unitlength}{1.5 true in}
\begin{picture}(1.75,1.5)(2.5,.875)
\thicklines
\put(3,1){\line(1,3){.5}}
\put(2.7,2.2){\mbox{$c_1^2=9\chi(\mathcal{O}_D)$}}
\put(3,1){\line(1,0){.1}}
\put(2.9,0.9){\mbox{$c_1^2=0$}}
\put(3.1,1){\line(1,1){.5}}
\put(3.45,1.2){\mbox{$c_1^2=2\chi(\mathcal{O}_D)-6$}}
\put(3.5,1.8){\mbox{$c_1^2=8\chi(\mathcal{O}_D)$}}

\color{red}
\put(3,1){\line(1,2){.5}}

\end{picture}
\caption{Constraints on invariants of surfaces of general 
                                  type}\label{fig:constraints}
\end{center}
\end{figure}

For curves, we had a simple invariant
(the genus), and rather complete knowledge about the
set of curves of genus $g$.
When $D$ is a surface, there are several invariants, including
the Euler number $e(D)=\chi_{\text{top}}(D)$ and
the degree of the canonical divisor $c_1^2=\#(K_D\cap K_D)$.
It is convenient to introduce
\begin{equation}
\chi(\mathcal{O}_D) = \frac{e(D)+c_1^2}{12},
\end{equation}
which is an integer, and to use $\chi(\mathcal{O}_D)$ in
place of $e(D)$. There are inequalities which constrain
these invariants to the region bounded by $c_1^2=9\chi(\mathcal{O}_D)$,
$c_1^2=0$, and $c_1^2=2\chi(\mathcal{O}_D)-6$, as illustrated
in Figure~\ref{fig:constraints}.  Surfaces are generally less numerous
above the central line $c_1^2=8\chi(\mathcal{O}_D)$ in the Figure 
than below it.\cite{BPV}

For each point on the graph, there are at most a finite number
of families, but it is not known how many, nor what are their
dimensions, etc.  See Persson\cite{geography} for a survey of what
{\it is}\/ known.

It has often been speculated that the number of families
of Calabi--Yau threefolds might be finite.  Certainly,
the vast array of possibilities of $D$, together with the
phenomenon of algebraic K3 surfaces for every $g\ge2$ casts
some doubt on this.  (However, as we shall see, the K3 surfaces
are in fact unified into a single family of hyper-K\"ahler
manifolds.)  Of course, many Calabi--Yau threefolds
have a wide variety of divisors $D$ on them, so there will
be much duplication. At the moment, it's hard to tell
whether the expectation that the number of families is
finite is reasonable or not.

\section*{Lecture IV: K3 Duality}

\section{Flat metrics on a two-torus}

In the previous lecture, we did not discuss the case of a torus $T^d$ in
any detail.
As we had earlier seen (lecture I), the set of flat metrics on $T^d$
admits a simple description:
\begin{equation}
\operatorname{Met}(T^d)= \SL(d,\mathbb Z)\backslash \mathbb R^\times \times 
\SL(d,\mathbb R)/\SO(d) 
\end{equation}
so the techniques of algebraic geometry were not needed.  However,
it is useful to see how algebraic geometry can be used to analyze this in 
the case $d=2$.

The conformal class of a metric on $T^2$ is equivalent to the
choice of complex structure.  Traditionally, one describes the
complex structure by representing $T^2$ as $\mathbb C/\langle1,\tau\rangle$,
i.e., $\mathbb C/\mathbb Z\oplus\mathbb Z\tau$, where the choice
of $\tau\in\upperhalf$ (the upper half plane) specifies the 
periodicity.  The one-cycles $\gamma_1$ and $\gamma_2$ which are
represented by curves in $\mathbb C$ joining the origin to $1$ and to
$\tau$, respectively,
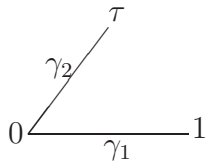
\[
\setlength{\unitlength}{0.0035in}
\begin{picture}(265,189)(100,585)
\thinlines
\put(240,760){\line(-3,-4){120}}
\put(120,600){\line( 1, 0){240}}
\put(240,770){\makebox(0,0)[lb]{\raisebox{0pt}[0pt][0pt]{$\tau$}}}
\put(365,595){\makebox(0,0)[lb]{\raisebox{0pt}[0pt][0pt]{$1$}}}
\put(90,587){\makebox(0,0)[lb]{\raisebox{0pt}[0pt][0pt]{$0$}}}
\put(232.5,570){\makebox(0,0)[lb]{\raisebox{0pt}[0pt][0pt]{$\gamma_1$}}}
\put(145,693){\makebox(0,0)[lb]{\raisebox{0pt}[0pt][0pt]{$\gamma_2$}}}
\end{picture}      
\]
give a basis of the first homology.
The torus has a holomorphic one-form, represented by
$dz$ in these coordinates, and the integrals over the generating
cycles $\gamma_1$, $\gamma_2$ give
$\int_{\gamma_1}dz=1$, $\int_{\gamma_2}dz=\tau$.  When we change basis
of $H_1(T^2,\mathbb Z)$ using $\SL(2,\mathbb Z)$, we get the standard
$\SL(2,\mathbb Z)$ action on the upper half plane.
(Note that $dz$ is only unique up to a constant multiple:
\begin{equation}
\int_{\gamma_1}A\cdot dz=A,\quad \int_{\gamma_2}A\cdot dz=A\tau
\end{equation}
so the truly invariant quantity is the ratio
$\int_{\gamma_2}dz/\int_{\gamma_1}dz=\tau$.)

Since $\upperhalf=\SL(2,\mathbb R)/\SO(2)$, we recover the description
of the moduli space $\SL(2,\mathbb Z)\backslash\upperhalf$.

To relate this to algebraic geometry, we need to study meromorphic
functions on $\mathbb C/\langle1,\tau\rangle$.  The basic such function
was studied by Weierstrass in the 19$^{\text{th}}$ century, called
the Weierstrass $\Wpee$-function:
\begin{equation}
\Wpee(z)=\frac1{z^2}+\sum_{(m,n)\ne(0,0)}
\left(\frac1{(z+m+n\tau)^2}-\frac1{(m+n\tau)^2}\right).
\end{equation}
This is doubly-periodic, with periods $1$, $\tau$, and has a double
pole at every lattice point, so descends to a meromorphic function
on $\mathbb C/\langle1,\tau\rangle$ with a double pole at the
origin.

Weierstrass found a remarkable relation which this function satisfies:
\begin{equation}
 (\Wpee'(z))^2=4(\Wpee(z))^3 -60G_4\Wpee(z)-140G_6 ,
\end{equation}
where
\begin{equation} 
G_4=\sum_{(m,n)\ne(0,0)}\frac1{(m+n\tau)^4}  , \quad
G_6=\sum_{(m,n)\ne(0,0)}\frac1{(m+n\tau)^6} . 
\end{equation}
If we define $y=\Wpee'(z)$, $x=\Wpee(z)$ and regard
$[1,x,y]$ as in $\mathbb P^2$, we find a cubic curve
\begin{equation}
y^2=4x^3-60G_4x-140G_6
\end{equation}
which is identical with $\mathbb C/\langle1,\tau\rangle$.
In homogeneous coordinates, this becomes
\begin{equation}
y^2w=4x^3-60G_4xw^2-140G_6w^3
\end{equation}
and the point $[0,0,1]$ was added representing the double pole
of $\Wpee(z)$.

This process can be reversed:  given a cubic curve $E$
\begin{equation}
y^2=4x^3-ax-b
\end{equation}
there is a non-vanishing holomorphic one-form given by the
residue of
\begin{equation}
\frac{2dx\wedge dy}{y^2-4x^3+ax+b}
\end{equation}
(i.e., by the adjunction formula, $K_E=(K_{\mathbb P^2}+3H)|_E
=-3H+3H|_E=0$).
The integration cycles are related to the branch cuts for
the function $y=\sqrt{4x^3-ax-b}$

\vskip.3in

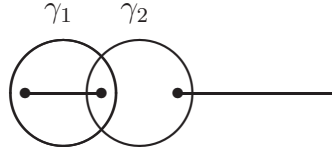
\begin{center}
\setlength{\unitlength}{1 true in}
\begin{picture}(3.4,.25)(1.4,1)
\thicklines
\put(3.0,1){\circle{1}}
\put(3.0,1){\circle{1}}
\put(3.4,1){\circle{1}}
\put(2.8,1){\line(1,0){.4}}
\put(2.8,1){\circle*{.05}}
\put(3.2,1){\circle*{.05}}
\put(3.6,1){\circle*{.05}}
\put(3.6,1){\line(1,0){.8}}
\put(2.9,1.4){\mbox{$\gamma_1$}}
\put(3.3,1.4){\mbox{$\gamma_2$}}

\end{picture}
\end{center}

\vskip.25in

\noindent
and we find periods given by {\it elliptic integrals}\/
\begin{equation} \label{eq:elliptic}
\int_{\gamma_j}\frac{dx}y=\int_{\gamma_j}\frac{dx}{\sqrt{4x^3-ax-b}}
\end{equation}
so that we recover
\begin{equation}
\tau= \int_{\gamma_2}\frac{dx}{\sqrt{4x^3-ax-b}} \ \bigg/  
\int_{\gamma_1}\frac{dx}{\sqrt{4x^3-ax-b}} .
\end{equation}
This is well-defined only up to $\SL(2,\mathbb Z)$.

\section{Kodaira's classification, and F-theory/heterotic duality}

We can now explain how Kodaira\cite{Kodaira} found the classification of
singular 
fibers in one-parameter elliptic fibrations, mentioned in lecture II.
If we consider families
\begin{equation}\label{eq:fam}
y^2=x^3+f(t)x+g(t)
\end{equation}
where now $f(t)$ and $g(t)$ are polynomials, we can try to classify
all possible behaviors near $t=0$ and the corresponding monodromy
on the periods.  Singularities occur when the cubic polynomial
has multiple roots, and that is measured by the {\it discriminant}:
\begin{equation}
\Delta(t)=4f(t)^3+27g(t)^2.
\end{equation}
Kodaira's analysis classifies possible monodromies in terms of
the divisibility properties $t^a\mathrel{\,|}f(t)$, $t^b\mathrel{\,|}g(t)$,
$t^c\mathrel{\,|}\Delta(t)$. 
It is a {\it local}\/ analysis in $t$.
For example, if $t$ does not divide both $f(t)$ and $g(t)$, and
$t^N\mathrel{\,|}\Delta(t)$, then we are in Kodaira's case $I_N$ (notation
as in Table~\ref{tab:Kodaira}).  The monodromy can be calculated by
analyzing how the elliptic integrals in Eq.~(\ref{eq:elliptic}) depend on
parameters. 

The order of zero of $\Delta(t)$ measures the deficit angle in the 
corresponding stringy cosmic string metric, with an angle of
$\pi m/6$ when 
$t^m \mathrel{\,|} \Delta(t)$, $t^{m+1} \not\mathrel{\,|} \Delta(t)$.

Thus, to find a global solution to the stringy cosmic string metric,
with total deficit angle $4\pi$, we need a total of $24$ zeros
of $\Delta(t)$.  This comes about precisely when $\deg f(t)=8$,
$\deg g(t)=12$.

Let us rewrite Eq.~(\ref{eq:fam}) in homogeneous form, as a surface $S$
in a $\mathbb P^2$ bundle over $\mathbb P^1$, with coordinates
$[w,x,y]$ on $\mathbb P^2$ and $[s,t]$ on $\mathbb P^1$.
It takes the form
\begin{equation}
y^2w=x^3+f_8(s,t)xw^2+g_{12}(s,t)w^3.
\end{equation}
Adapting the adjunction formula to this situation shows that
$K_S=0$, i.e., a stringy cosmic string solution which can be
used to build an F-theory model corresponds to a K3 surface when
the $\tau$-function is realized by elliptic curves.
Thus, we expect that when F-theory models in eight dimensions
are compactified on a circle, the result is M-theory on  a K3 surface.
(We have already encountered this possibility in our remarks
about T-dualizing type I on $T^3$.)  This leads to the first of
the K3 duality statements: {\it F-theory from K3 is dual to the
heterotic string on $T^2$}.

Going up one dimension, we can ask if there is a K3 interpretation
for the nine-dimensional heterotic string models.  Such an interpretation
has been proposed\cite{MVII} as a limit in which the
$S^2$ stretches to a long cylinder.  A more precise realization of
this picture has recently been worked out by 
Cachazo and Vafa\cite{CachazoVafa}---it 
involves {\it real}\/ K3 surfaces, i.e., restricting the coefficients
in Eq.~(\ref{eq:fam}) to real numbers and searching for real solutions.

\section{M-theory/heterotic duality}

To find K3 duality statements in lower dimension, we need to study
the Ricci-flat metrics on K3, or more generally on hyper-K\"ahler
manifolds.\cite{beauville:varietes}

Let $X^{4d}$ be a hyper-K\"ahler manifold.  A somewhat abstract 
description of the Ricci-flat metrics on $X$ goes like
this:\cite{beauville:remarks} 
$H^2(X,\mathbb R)$ has a natural inner product
\begin{equation}
\begin{aligned}
H^2(X,\mathbb R)\times H^2(X,\mathbb R) &\to \mathbb R \\
(\alpha,\beta) &\mapsto \int_X\alpha\wedge\beta\wedge\omega^{2d-2}
\end{aligned}
\end{equation}
(which only changes by a scale factor if the K\"ahler form
$\omega$ is changed).
The {\it signature}\/ of this inner product is $(3,k)$,
where $b_2(X)=3+k$ depends on $X$.  (For a K3 surface, $k=19$.)

Each Ricci-flat metric on $X$ determines a positive three-plane
in $H^2(X,\mathbb R)$, the ``self-dual'' harmonic two-forms
(with respect to the inner product above).  The space of all
positive three-planes is
\begin{equation}
\Gamma\backslash\OO(3,k)/\OO(3)\times\OO(k) ,
\end{equation}
where we have taken the quotient by $\Gamma$ which comes from
the diffeomorphism group of $X$.  So there is a map
\begin{equation}
\{\text{Ricci-flat metrics}\}\to 
\left(\Gamma\backslash\OO(3,k)/\OO(3)\times\OO(k)\right)\times\mathbb R^+
\end{equation}
(with the last $\mathbb R^+$ representing the volume).
However, it is very difficult to directly determine whether this
map is one-to-one or onto.  It {\it is}\/ possible to show that
locally, the map is one-to-one and onto, i.e., small variations of
the Ricci-flat metric are accurately reflected by small variations
of the positive three-plane and the volume.

To give a more concrete interpretation of this space, and relate it
to Yau's theorem and algebraic geometry, we must choose a complex
structure compatible with the Ricci-flat metric.  Rather than study
the totality of all such complex structures,
we will select an element 
$\lambda\in H^2(X,\mathbb Z)$ for which
$\int_X\lambda\wedge\lambda\wedge\omega^{2d-2}>0$,
and choose the complex structure so that $\lambda$ becomes type
$(1,1)$.  To do that, if $\Pi$ denotes our three-plane, then
$\lambda^\perp\cap\Pi$ is a two-plane; we let $\omega_1$,
$\omega_2$ be a basis of that two-plane and use the complex
structure for which $\omega_1+i\omega_2$ is holomorphic
(and $\omega_3\in\omega_1^\perp\cap\omega_2^\perp\cap\Pi$
is a K\"ahler form).

The advantage of this choice is that $\lambda$ will now be an
algebraic class, since it is integral and type $(1,1)$,
and for general moduli some multiple $m\lambda$ will be a
{\it very ample class}.
Thus, we can study hyper-K\"ahler manifolds which are embedded
in $\mathbb P^n$, with some specific type of embedding, and
capture all the information we need about metrics on the space
in that context.\,\footnote{Note that we are getting complete information
only about the open subset in the moduli space where $m\lambda$ is very
ample; if we vary $\lambda$ and/or $m$, we can change this open set.  The
full moduli space will be the union of open sets of this kind.}

With respect to our chosen complex structure (which we could
describe algebraically, as in the case of all the algebraic
K3 surfaces in lecture III), the set of all K\"ahler classes will
correspond to the set of Ricci-flat metrics, by Calabi's and
Yau's theorems.  However, unlike the Calabi--Yau case we cannot
give a purely algebraic description of the K\"ahler classes.
The defining conditions we need, for possible K\"ahler classes
$\kappa$ (with respect to our complex structure) are:
\begin{enumerate}
\item\label{en:one} $\kappa\cdot\omega_1=0$
\item\label{en:two} $\kappa\cdot\omega_2=0$
\item\label{en:three} $\kappa\cdot\kappa>0$
\item\label{en:four} $\int\kappa^\ell\alpha>0$ for every $\alpha$ representing
the cohomology class of a complex submanifold of $X$ of complex codimension
$\ell$.
\end{enumerate}
Conditions \ref{en:one} and \ref{en:two} restrict
$\kappa$ to a space of dimension $b_2-2=k+1$;
conditions \ref{en:three} and \ref{en:four}
select an open subset and do not affect the dimension.

Moreover, it is known that the set of algebraic deformations of $X$ has
complex dimension $k$ (and the set of {\it all}\/ complex deformations
has complex dimension $k+1$).  As a check, then, counting real parameters
for our metric we find $2k$ from the algebraic deformations of
complex structure and $k+1$ from the compatible K\"ahler metrics,
for a total of
\begin{equation}
3k+1=\dim_{\mathbb R}\left(\mathbb
R^+\times\OO(3,k)/(\OO(3)\times\OO(k))\right).
\end{equation}

Focusing on algebraic deformations allows us to apply some of the machinery
from algebraic geometry to study the variations of complex structure.
In the case of K3 surfaces, there is a ``Torelli theorem'' for this moduli
space\,\footnote{The Torelli theorem for K3 surfaces has a long history.  An
account of the original theorem can be found in the ``S\'eminaire
Palaiseau.''\cite{Palaiseau}  The version stated here was proved directly
by Anderson\cite{anderson}, and can also be extended to the singular
set $Z$.\cite{Mor:Katata,Kobayashi-Todorov}} which tells us the precise answer:
the set of metrics 
is given by
\begin{equation}
\OO(3,19;\mathbb Z)\backslash \left(\OO(3,19)/\OO(3)\times\OO(19) - Z
\right)\times\mathbb R^+ ,
\end{equation}
where 
\begin{equation}
Z
{=}
\{\text{$\Pi$ 
such that 
$\Pi\cdot e=0$ for some
$e\in H^2(X,\mathbb Z)$ with $e\cdot e=-2$}\}.
\end{equation}
(Such planes $\Pi\in Z$ {\it cannot}\/ correspond to metric,
since $\pm e$ represents the class of a rational curve on $X$,
which would have vanishing area in such a ``metric.'')

As we saw when discussing the T-duals of type I on $T^3$, we are
expecting a duality between M-theory on K3 and the heterotic string
on $T^3$.  For M-theory compactification, the scalars in the effective
theory will be given by the set of Ricci-flat metrics on K3
as well as by harmonic three-forms.  But K3 manifolds have no
harmonic three-forms, so the entire M=theory moduli space is
\begin{equation}
\OO(3,19;\mathbb Z)\backslash \left(\OO(3,19)/\OO(3)\times\OO(19) - Z
\right)\times\mathbb R^+ .
\end{equation}
This {\it agrees}\/ with the heterotic string on $T^3$,
except for the phenomenon of the subset $Z$.

What is the interpretation of $Z$?  As described above, the problem
we encounter along $Z$ is that a holomorphic $S^2$ is shrinking to
zero area.  However, M-theory contains more than supergravity,
and in particular we can wrap the M-theory membrane around this
$S^2$.  The mass of the corresponding state is proportional to the
area, so along $Z$ we will find new massless states, corresponding
to the $e$'s with $e\cdot\Pi=0$.

On the heterotic side, $Z$ is the locus along which the gauge
symmetry becomes non-abelian.  Now we are finding the source of
non-abelian gauge symmetry in M-theory: massless multiplets
from shrinking $S^2$'s, which are {\it vectors}\/ (new gauge fields)
in the spectrum of the effective theory.

The geometry of configurations of $e$'s is quite pretty.
We need to study collections of holomorphic $S^2$'s 
(i.e., $\mathbb P^1$'s) embedded in $X$ whose intersection matrix
$(e_i\cdot e_j)$ is negative definite.  The restriction to
negative definite matrices arises because these $e$'s lie
in $\Pi^\perp$, which is a negative definite space of dimension
$19$.
The entries in the intersection matrix are $e_i\cdot e_i=-2$
(from the adjunction formula again, since $g=0$), and
$e_i\cdot e_j=0 \text{ or } 1$ when $i\ne j$.
($e_i\cdot e_j\ge2\implies (e_i+e_j)^2\ge0$, a contradiction.)

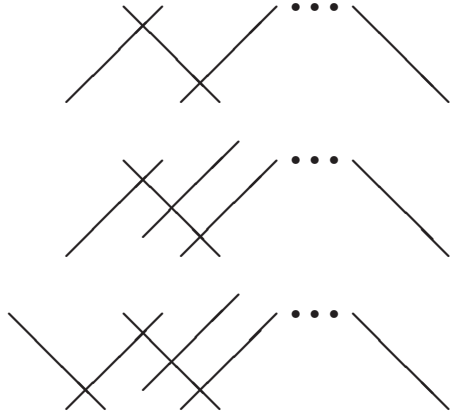
\begin{figure}[h,t]

\vskip.3in

\begin{center}
\setlength{\unitlength}{1 true in}
\begin{picture}(3.4,.25)(1.4,1)
\thicklines
\put(2.4,1){\line(1,1){.5}}
\put(2.7,1.5){\line(1,-1){.5}}
\put(3.0,1){\line(1,1){.5}}
\put(3.6,1.5){\circle*{.04}}
\put(3.7,1.5){\circle*{.04}}
\put(3.8,1.5){\circle*{.04}}
\put(3.9,1.5){\line(1,-1){.5}}
\end{picture}
\end{center}

\vskip.5in

\begin{center}
\setlength{\unitlength}{1 true in}
\begin{picture}(3.4,.25)(1.4,1)
\thicklines
\put(2.4,1){\line(1,1){.5}}
\put(2.7,1.5){\line(1,-1){.5}}
\put(2.8,1.1){\line(1,1){.5}}
\put(3.0,1){\line(1,1){.5}}
\put(3.6,1.5){\circle*{.04}}
\put(3.7,1.5){\circle*{.04}}
\put(3.8,1.5){\circle*{.04}}
\put(3.9,1.5){\line(1,-1){.5}}
\end{picture}
\end{center}

\vskip.5in

\begin{center}
\setlength{\unitlength}{1 true in}
\begin{picture}(3.4,.25)(1.4,1)
\thicklines
\put(2.1,1.5){\line(1,-1){.5}}
\put(2.4,1){\line(1,1){.5}}
\put(2.7,1.5){\line(1,-1){.5}}
\put(2.8,1.1){\line(1,1){.5}}
\put(3.0,1){\line(1,1){.5}}
\put(3.6,1.5){\circle*{.04}}
\put(3.7,1.5){\circle*{.04}}
\put(3.8,1.5){\circle*{.04}}
\put(3.9,1.5){\line(1,-1){.5}}
\end{picture}

\caption{Dynkin diagrams for ADE groups}\label{fig:Dynkin}
\end{center}
\end{figure}

This is precisely the same combinatorial problem as the one
which classified simply-laced Dynkin diagrams, and the answer
is the same, illustrated in Figure~\ref{fig:Dynkin}.
In the Figure,
we have drawn the holomorphic curves---each of self-intersection
$-2$---and indicated which ones meet.

In M-theory, we will associate gauge algebras $\su(n)$, $\so(2n)$,
$\e_6$, $\e_7$, $\e_8$ to these cases.  In fact, the set of classes
$e=\sum m_ie_i$ for which $e\cdot e=-2$ precisely correspond
to the roots in the Lie algebra, and we will get a gauge boson
from the M-theory membrane wrapped around each such class.

The picture in algebraic geometry of these singular spaces is
fairly benign; the coefficients of the equations  have been
tuned to special values at which such singularities appear.
The singularities can be removed either by varying the complex
structure, or by ``blowing up'' the algebraic variety which
effectively increases the area assigned to $e_j$.

One description of these singularities is that they are precisely
the orbifolds $\mathbb C^2/\Gamma$ where $\Gamma\subseteq \SU(2)$
is a finite subgroup acting without fixed points away from
the origin.  The dictionary  was already given in Table~\ref{tab:ADE}.
Note that one lesson from this analysis is that M-theory on an orbifold
is only well-behaved when the non-abelian gauge theory coming from
the new massless vectors is included.

\section{Type IIA/heterotic duality}

When we extend this analysis to type IIA on K3, we find that the
orbifold points will be accompanied by a choice of $B$-field value.
When the $B$-field is zero, the perturbative string is singular,
and the new massless states from wrapped D$2$-branes must be
included.  However, the theory is perturbatively non-singular
with $B$-fields turned on; in particular, Aspinwall\cite{AspEnhanced}
checked that 
the $B$-field is non-zero at the orbifold conformal field theories,
so string theory is indeed nonsingular (even at the perturbative
level) at such points.

The summary of our K3 dualities is:
\smallskip

\centerline{het on $T^2$ $\leftrightarrow$ F-theory from K3}

\smallskip
\centerline{het on $T^3$ $\leftrightarrow$ M-theory on K3}

\smallskip
\noindent
and, as we have just briefly indicated, this extends to
\smallskip

\centerline{het on $T^4$ $\leftrightarrow$ IIA on K3}

\smallskip
\noindent
in a similar manner.

The final duality is bolstered by a construction of a soliton of the type
IIA string theory compactified on K3,\cite{HS} which in the appropriate
limit exhibits the characteristics of a heterotic string.

\section*{Acknowledgments}

I would like to thank the
Institute for Advanced Study, Princeton,  and the Institute for Theoretical
Physics, Santa Barbara, for hospitality during various stages of writing
these notes.
This work was partially supported by 
National Science Foundation grants DMS-9401447, DMS-0074072, and
PHY-07949, and by the Institute for Advanced Study.

\section*{References}
\ifx\undefined\bysame
\newcommand{\bysame}{\leavevmode\hbox to3em{\hrulefill}\,}
\fi

\end{document}